\pdfoutput=1
\documentclass[11pt,a4paper]{article}

\usepackage{jheppub}
\usepackage{amsmath,amsfonts,amssymb,graphics}
\usepackage[normalem]{ulem}
\usepackage{graphicx}
\usepackage{mathrsfs}
\usepackage{bbm}
\usepackage{pifont}
\usepackage{braket}
\usepackage{footmisc}
\usepackage[dvipsnames]{xcolor}
\usepackage{mathtools}
\usepackage{xspace}
\usepackage{booktabs}
\usepackage{siunitx}
\usepackage{tikz}
\usetikzlibrary{positioning,shapes,arrows}
\usetikzlibrary{decorations.pathmorphing}
\usetikzlibrary{decorations.markings}
\usepackage{standalone}
\usepackage{csquotes}

\title{ Freeze-In of  radiative keV-scale neutrino dark matter from a new \boldmath $\text{U}(1)_\text{B-L}$}
\author{Maximilian Berbig}
\affiliation{Bethe Center for Theoretical Physics und Physikalisches 
Institut der Universit\"at Bonn,\\
Nussallee 12, Bonn, Germany}
\emailAdd{berbig@physik.uni-bonn.de}

\abstract{We extend the Dirac Scotogenic model with the aim of realizing  neutrino masses together with the mass of a keV-scale  dark matter (DM) candidate via the same one-loop topology.  Two of the Standard Model (SM) neutrinos become massive Dirac fermions while the third one remains massless.
Our particle content is motivated by an anomaly free $\text{U}(1)_\text{B-L}$ gauge symmetry with exotic irrational charges and we need to enforce an additional $\mathcal{Z}_5$ symmetry. The dark matter candidate does not mix with the active neutrinos and does not have any decay modes to SM particles.
DM is produced together with dark radiation in the form of right handed neutrinos via out of equilibrium annihilations of the SM fermions mediated by the heavy B-L gauge boson. In order to avoid DM over-production from Higgs decays and to comply with Lyman-$\alpha$ bounds  we work  in a low temperature reheating scenario with $\SI{4}{\mega\electronvolt}\lesssim T_\text{RH}\lesssim \SI{5}{\giga\electronvolt}$.   Our setup predicts a contribution to $\Delta N_\text{eff.}$ that decreases for larger DM masses  and is below the sensitivity of upcoming precision measurements such as CMB-S4. A future observation of a signal with $\Delta N_\text{eff.}\gtrsim 0.012$ would exclude our scenario. We further sketch how inflation, reheating and Affleck-Dine baryogenesis can also be potentially realized in this unified framework.}
\keywords{neutrino masses, dark matter, freeze-in, dark radiation}

\begin{document}
\maketitle

\section{Introduction}
\noindent In most models of the Scotogenic variety one uses the lightest stable particle from the loop diagram for the neutrino masses as a DM candidate. For the original scotogenic model  \cite{Ma:2006km,Ma:2006fn,Kubo:2006yx,Hambye:2006zn} this is either the lightest neutral component of the inert scalar doublet $\eta$ or the lightest  sterile neutrino produced as thermal WIMPS. In the Dirac version of the model \cite{Gu:2007ug,Farzan:2012sa} the DM can either the lightest neutral component of $\eta$ or the singlet $\sigma$ or the lightest vector-like neutrino. Later it was realized that keV scale FIMP DM is also possible in the scotogenic picture \cite{Molinaro:2014lfa}, but no mechanism was proposed for why this particular sterile neutrino is so much lighter than the other two.
Reference \cite{Dedes:2017shn} analyzed a model based on the DFSZ axion scenario \cite{Zhitnitsky:1980tq,Dine:1981rt}, where a one loop diagram with vector-like fermions generates the keV-scale Majorana masses for a DM candidate. The authors of \cite{Borah:2013waa,Adhikari:2014nea,Adhikari:2015woo} showed that it is possible to construct models in which both the active and the sterile neutrino masses are obtained from loop diagrams.
Recently a loop based extension of the seesaw scenario \cite{Gell-Mann:1979vob,Sawada:1979dis,PhysRevLett.44.912}  with keV to GeV scale Majorana dark matter was put forth in \cite{Ma:2021eko}, where two different scalar couplings were responsible for the mass generation and production from out of equilibrium Higgs decays. Unlike previous constructions we focus on Dirac neutrinos. We choose  an abelian gauge symmetry as the guiding principle for building our model. After reviewing the Dirac Scotogenic model in section \ref{sec:review} we introduce our mechanism for generating the DM mass via a dimension five operator that resembles the Weinberg operator \cite{WeinbergOperator}  in \ref{sec:ext}. 
In \ref{sec:higgs} we find that producing such a  dark matter from out of equilibrium Higgs decays is not compatible with Lyman-$\alpha$ bounds on the DM mass.
Section \ref{sec:gauge1} demonstrates that the gauge symmetry is crucial for producing the correct amount of DM in the freeze-in scenario. We compute the minuscule amount of dark radiation produced by a similar freeze-in process in section \ref{sec:darkRad}. The necessary cosmic history can be realized in an inflationary context as explained in section \ref{sec:infl}. We close by illustrating how our set-up can potentially realize Affleck-Dine baryogenesis \cite{AFFLECK1985361} in section \ref{sec:baryo}.

\section{The model}
\subsection{The Dirac Scotogenic model}\label{sec:review}

\begin{figure}[t]
 \centering
  \tikzset{
  blackline/.style={thin, draw=black, postaction={decorate},
    decoration={markings, mark=at position 0.6 with {\arrow[black]{triangle 45}}}},
    blueline/.style={thin, draw=blue, postaction={decorate},
    decoration={markings, mark=at position 0.6 with {\arrow[blue]{triangle 45}}}},
    redline/.style={thin, draw=red, postaction={decorate},
    decoration={markings, mark=at position 0.6 with {\arrow[red]{triangle 45}}}},
    greenline/.style={thin, draw=green, postaction={decorate},
    decoration={markings, mark=at position 0.6 with {\arrow[green]{triangle 45}}}},    
    graydashedarrow/.style={dashed, draw=gray,  postaction={decorate},
    decoration={markings, mark=at position 0.6 with {\arrow[gray]{triangle 45}}}},
    graydashed/.style={dashed, draw=gray, postaction={decorate},decoration={markings}},
    photon/.style={decorate, draw=red,
    decoration={coil,amplitude=12pt, aspect=0}},
    reddashed/.style={thick, dashed, draw=red, postaction={decorate},
    decoration={markings}},
    photon/.style={decorate, draw=red,
    decoration={coil,amplitude=12pt, aspect=0}},
  gluon/.style={dashed, decorate, draw=black,
    decoration={coil, segment length=5pt, amplitude=8pt}}
  line/.style={thick, draw=black, postaction={decorate},
    decoration={markings}}
}

\begin{tikzpicture}[node distance=1cm and 1cm]

\coordinate[label = below: $L$] (start1);
\coordinate[right=1.5cm of start1] (vertex1);
\coordinate[right=1.5cm of vertex1] (vertex2);
\coordinate[right=1.5cm of vertex2] (vertex3);
\coordinate[above=1.5cm of vertex2] (vertex4);
\coordinate[above=1.2cm of vertex4, label = above: $\braket{H}$] (vev1);
\coordinate[right=1.5cm of vertex3,label = below: $\nu_R$] (end1);

\coordinate[above=1.2cm of vertex1, label= above: $\eta$] (label1);
\coordinate[above=1.2cm of vertex3, label= above: $\sigma$] (label2);
\coordinate[below=0.3cm of vertex2] (middle1);
\coordinate[left=0.75cm of middle1, label = below: $N_R$] (label3);
\coordinate[right=0.75cm of middle1, label = below: $N_L$] (label4);

\fill (vertex1) circle (2pt);
\fill (vertex2) circle (2pt);
\fill (vertex3) circle (2pt);
\fill[gray] (vertex4) circle (2pt);

\draw[blackline] (start1)   -- (vertex1);
\draw[blackline] (vertex1)   -- (vertex2);
\draw[blackline] (vertex2) -- (vertex3);
\draw[blackline] (vertex3) -- (end1);
\draw[graydashed] (vertex4) -- (vev1);

\draw[graydashedarrow] (vertex3) arc[start angle=0, end angle=90, radius=1.5];
\draw[graydashedarrow] (vertex4) arc[start angle=-270, end angle=-180, radius=1.5];

 
\coordinate[right= 3cm of end1, label = below: $\chi_L$] (start2);
\coordinate[right=1.5cm of start2] (vertex21);
\coordinate[right=1.5cm of vertex21] (vertex22);
\coordinate[right=1.5cm of vertex22] (vertex23);
\coordinate[above=1.5cm of vertex22] (vertex24);
\coordinate[above=1.2cm of vertex24, label = above: $\braket{H}$] (vev2);
\coordinate[right=1.5cm of vertex23,label = below: $\chi_R$] (end2);
\coordinate[below=1.2cm of vertex22, label = below: $\braket{H}$] (vev3);

\coordinate[above=1.2cm of vertex21, label= above: $\eta$] (label21);
\coordinate[above=1.2cm of vertex23, label= above: $\sigma$] (label22);
\coordinate[below=0.3cm of vertex22] (middle2);
\coordinate[left=0.75cm of middle2, label = below: $D_R$] (label23);
\coordinate[right=0.75cm of middle2, label = below: $S_L$] (label24);

\fill (vertex21) circle (2pt);
\fill (vertex22) circle (2pt);
\fill (vertex23) circle (2pt);
\fill[gray] (vertex24) circle (2pt);

\draw[blackline] (start2)   -- (vertex21);
\draw[blackline] (vertex21)   -- (vertex22);
\draw[blackline] (vertex22) -- (vertex23);
\draw[blackline] (vertex23) -- (end2);
\draw[graydashed] (vertex24) -- (vev2);
\draw[graydashed] (vertex22) -- (vev3);

\draw[graydashedarrow] (vertex21) arc[start angle=180, end angle=90, radius=1.5];
\draw[graydashedarrow] (vertex24) arc[start angle=90, end angle=0, radius=1.5];

\end{tikzpicture}
  \caption{Feynman diagrams in the gauge basis responsible for the creation of the neutrino and dark matter ($\chi$) Dirac masses at the one loop level.}
  \label{fig:mass-nu-DM}
\end{figure}
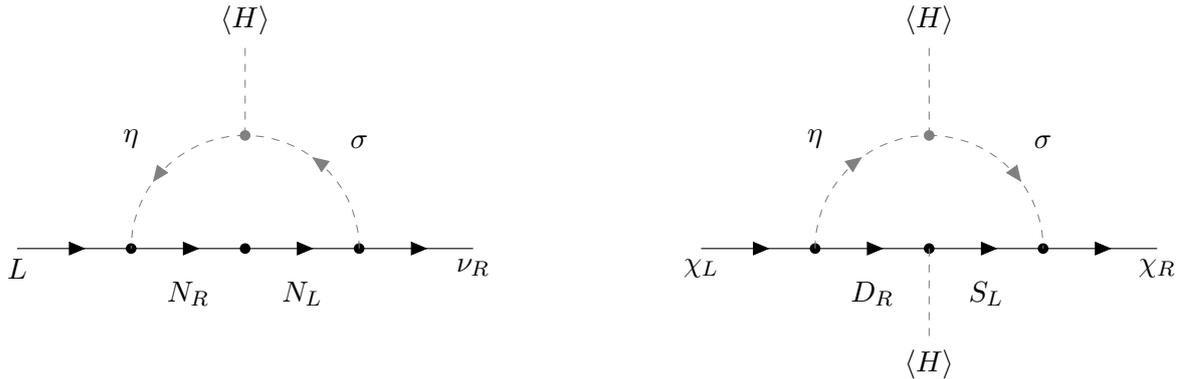

\noindent 
Let us begin by reviewing the most salient features of the Scotogenic Model for Dirac neutrinos \cite{Gu:2007ug,Farzan:2012sa}. The goal is to generate the first diagram in figure \ref{fig:mass-nu-DM}. We follow the treatment of \cite{Ma:2019coj}, where a $\text{U}(1)$ symmetry is imposed on the fermionic sector that gets  softly broken by the following trilinear term in the scalar potential
\begin{equation}\label{eq:tril}
   V(H,\eta,\sigma) \supset\quad  \frac{\kappa}{\sqrt{2}} \;\left(\eta^\dagger H \sigma\;  +  \; \text{h.c.}\right),
\end{equation}
where $\kappa$ is a dimensionful parameter of mass dimension one. Here $H$ is the SM Higgs and $\eta, \sigma$ are  inert doublet and singlet scalars with charges under the new symmetry. All particles and charges can be found in table \ref{tab:charges-reps}. We start from $\text{U}(1)_\text{B-L}$ and  assign conventional B-L charges -1 to $L$ and $e_R$, whereas the right handed neutrinos $\nu_R$ have the charge $Q_1\neq-1$ so that the tree level mass term $\overline{L}\epsilon H^\dagger \nu_R$ is forbidden by the symmetry. Here $\epsilon = i\sigma_2$ denotes the anti-symmetric tensor in two dimensions. 

\begin{table}[!t]
\centering
 \begin{tabular}{|c||c|c|c|c||c|} 
 \hline
  field&     $\text{SU}(2)_\text{L}$ & $\text{U}(1)_\text{Y}$ & $\text{U}(1)_\text{B-L}$ &$\mathcal{Z}_5$& generations\\
 \hline
 \hline 
    $L$  & 2& $-1/2$ & $-1$& $-4$& 3 \\
    $e_R$  & 1 & $-1$& $-1$& 1&3\\
 \hline
    $H$& 2& $1/2$& 0 & 0&1 \\
 \hline
 \hline
    $\nu_R$& 1 & 0 & $-2$ & 1&2\\
    $N_L$& 1 & 0 & $-3$ & $-3$&2\\
    $N_R$& 1 & 0 & $-3$ & 2&2\\
\hline
    $\chi_L$& 1 & 0 & $Q_4$ & 0&1\\
    $\chi_R$& 1 & 0 & $Q_3$ & 0&1\\
    $D_L$& 2 & $-1/2$ & $1+Q_3$ & $-1$&1\\
    $D_R$& 2 & $-1/2$ & $1+Q_3$ & 4&1\\
    $S_L$& 1 & 0 & $1+Q_3$ & $-1$&1\\
    $S_R$& 1 & 0 & $1+Q_3$ & 4&1\\
\hline
    $\eta$ & 2 &  $1/2$& $-2$ & $-4$&1\\
    $\sigma$ & 1 & 0 & $-1$ & 1&1\\
    $ \phi$ & 1 & 0 & 1& 0& 1 \\
\hline 
\end{tabular}
\caption{Charges and representations for all particles participating in the neutrino or dark matter mass generation. The integers $n$ in the fifth column are an abbreviation for $\omega^n$, where $\omega = e^\frac{2 i\pi}{5}$.}
\label{tab:charges-reps}
\end{table}

\noindent To generate this operator at loop level requires a soft $\text{U}(1)_\text{B-L}$ breaking by $1+Q_1$ units. Since we assume that $H$ is uncharged under the new group, this means that the term $\kappa\;\eta^\dagger H \sigma$ has to have the same total charge $Q_\eta-Q_\sigma =  1+Q_1$. This soft breaking can be UV completed  by considering  the vev $\kappa = \lambda_\text{IV} v_\phi$ of another singlet scalar $\phi$ with charge $-1-Q_1$, as will be shown in section \ref{sec:UV}. On the fermionic side we introduce two generations of  vector-like pairs of SM singlets $(N_L,N_R)$ with B-L charge $Q_N$
\begin{equation}
    \mathcal{L}  \supset -Y_{LN} \;\overline{L} \epsilon \eta^\dagger N_R - Y_{NR}\; \overline{N_L}\sigma \nu_R - M_N \overline{N_L}N_R + \text{h.c.}.
\end{equation}
In order to forbid a Dirac mass with $L$
and $\nu_R$ we have to require that $Q_N \neq \pm 1, \pm Q_1$. We also need to forbid the following  operators \cite{Ma:2019coj}:
\begin{itemize}
    \item $\overline{N^c_L},N_L$ and $\overline{N^c_R} N_R$ with $2 Q_N$
    \item $\overline{\nu_R^c} \nu_R$ with $2 Q_1$
    \item $\overline{N_L}\nu_R, \overline{N_R^c} \nu_R$ with $-Q_N+Q_1, Q_N+Q_1$
     \item $\left(H^\dagger \eta\right)\left(H^\dagger \eta\right)$  with $2 Q_\eta$ together with $\overline{N^c_R},N_R$ would create $\overline{\nu^c_L}\nu_L$ at loop level \cite{Ma:2006km}
    \item $\sigma \sigma$ with $2 Q_\sigma$ together with $\overline{N^c_L},N_L$ would create $\overline{\nu^c_R}\nu_R$ at loop level  \cite{Ma:2021eko}
\end{itemize}

\noindent All of the above combinations of charges need to be non-zero and not divisible by $|1+Q_1|$. If they were divisible by the only source of soft breaking, then an integer number of insertions of the trilinear scalar coupling in  some loop diagram can generate the unwanted mass term. Once we know $Q_1$ we can fix all the other charges of the model. We will use the criterion of anomaly freedom to determine the rest of the particle spectrum and to find $Q_1$ in the next section. Before we do let us continue with our short review of the Dirac Scotogenic model: The active Dirac neutrino mass arises due to the first diagram in \ref{fig:mass-nu-DM} and depends on the mass mixing in the scalar sector:
\begin{eqnarray}
    \mathcal{L}\supset &-& m_\sigma^2 \left|\sigma\right|^2 - m_\eta^2 \left|\eta\right|^2 - 
    \frac{\kappa}{\sqrt{2}}\left( \;\eta^\dagger H \sigma+\; \text{h.c.}\right)\\
    &-& \lambda_\eta \left(\eta^\dagger \eta\right)^2 - \lambda_\sigma  \left|\sigma\right|^4\\
    &-& \lambda_{H\eta\;1} \left(H^\dagger H\right) \left(\eta^\dagger \eta\right) - \lambda_{H\eta\;2}  \left(H^\dagger \eta\right) \left(\eta^\dagger H\right) \\
    &-& \lambda_{H\sigma}  \left(H^\dagger H\right) \left|\sigma\right|^2 \label{eq:Hsigma}
\end{eqnarray}
After we expand all the fields into their components
\begin{equation}\label{eq:reps}
    H = \begin{pmatrix}h^+\\ \frac{h_R+v_H+i h_I}{\sqrt{2}} \end{pmatrix}, \; \eta = \begin{pmatrix}\eta^+\\ \frac{\eta_R^0 + i \eta_I^0}{\sqrt{2}} \end{pmatrix}, \; \sigma = \frac{\sigma_R^0 + i \sigma_I^0}{\sqrt{2}}
\end{equation}
and in the absence of $CP$-violation there is no mass mixing between the $CP$-even (subscript $R)$ and odd bosons  (subscript $I$). We set $m_\eta^2, m_\sigma^2>0$ in order to have an inert doublet and singlet.
The real and imaginary components only mix among each other. The mass matrix after EWSB reads
\begin{equation}
  \left(\eta_R^0,\sigma_R^0\right) \cdot \begin{pmatrix}\tilde{m}_\eta^2 & \frac{\kappa v_H}{2} \\ \frac{\kappa v_H}{2} & \tilde{m}_\sigma^2 \end{pmatrix}\cdot \begin{pmatrix}\eta_R^0\\ \sigma_R^0\end{pmatrix},
\end{equation}
and the same holds for the $CP$-odd fields, where 
\begin{eqnarray}\label{eq:mass}
    \tilde{m}_\eta^2 \equiv  m_\eta^2 + \left(\lambda_{H\eta\;1}+\lambda_{H\eta\;2}\right)v_H^2, \quad \text{and} \quad  \tilde{m}_\sigma^2 \equiv m_\sigma^2 + \lambda_{H\sigma}v_H^2.
\end{eqnarray}
We find two mass eigenstates in each case with the masses 
\begin{equation}
    m_{1,2}^2 = \frac{1}{2}\left( \tilde{m}_\eta^2+\tilde{m}_\sigma^2\pm \sqrt{\left( \tilde{m}_\eta^2-\tilde{m}_\sigma^2\right)^2+\kappa^2v_H^2}\right)
\end{equation}
and the mass eigenstates read
\begin{equation}
    \begin{pmatrix} \eta_R^0 \\ \sigma_R^0\end{pmatrix} = \begin{pmatrix} \cos(\alpha)& \sin(\alpha) \\-\sin(\alpha) & \cos(\alpha) \end{pmatrix}\begin{pmatrix}S_1\\ S_2\end{pmatrix}
    , \quad  
    \begin{pmatrix} \eta_I^0 \\ \sigma_I^0\end{pmatrix} = \begin{pmatrix} \cos(\alpha)& \sin(\alpha) \\-\sin(\alpha) & \cos(\alpha) \end{pmatrix}\begin{pmatrix}A_1\\ A_2\end{pmatrix}.
\end{equation}
The mixing angle is given in terms of the model parameters as 
\begin{equation}\label{eq:sc-mixing}
    \sin(2\alpha) = \frac{\kappa v_H}{2 \Delta m_S^2}, \quad \text{with} \quad \Delta m_S^2 \equiv \frac{m_1^2-m_2^2}{2}.
\end{equation}
Four diagrams contribute to the active neutrino masses: one mediated by each of the scalars $S_{1,2}$ and $A_{1,2}$. Since $S_1$ and $A_1$ ($S_2$ and $A_2$) are mass degenerate there are only two distinct types of diagrams: two for  heavier scalars of mass $m_1$ and two for the ones with $m_2$. Due to the mixing there will be a relative sign between these two \enquote{generations} of scalars. This difference cancels out the divergent part leaving us with a finite mass matrix \cite{Escribano:2020iqq}
\begin{equation}
    \left(m_\nu\right)_{ij} = -\frac{\sin(2\alpha)}{32\pi^2}\sum_{k=1}^2 \left(Y_{LN}\right)_{ik} \left(Y_{NR}\right)_{kj} M_N^{(k)} \left[ 
    \frac{m_2^2\; \text{Log}\left(\frac{m_2^2}{M_N^{(k)\;2}}\right)}{m_2^2-M_N^{(k)\;2}}
    - \frac{m_1^2\; \text{Log}\left(\frac{m_1^2}{M_N^{(k)\;2}}\right)}{m_1^2-M_N^{(k)\;2}} \right],
\end{equation}
where $M_N^{(k)}$ is the mass of the $k$-th heavy neutrino. To get a more insightful expression we work in the radiative seesaw limit \cite{Ma:2006km}
\begin{equation}
   M_N^{(k)\; 2} \gg m_0^2\equiv\frac{m_1^2+m_2^2}{2} \gg \Delta m_S^2.
\end{equation}
After substituting in the mixing angle from \eqref{eq:sc-mixing} we find
\begin{equation}\label{eq:act-mass}
    \left(m_\nu\right)_{ij} =\sum_{k=1}^2 \frac{ \left(Y_{LN}\right)_{ik} \left(Y_{NR}\right)_{kj} }{32\pi^2} \frac{\kappa v_H}{M_N^{(k)}}\left(\text{Log}\left(\frac{M_N^{(k)\;2}}{m_0^2}\right)-1\right),
\end{equation}
where the dependence on the soft symmetry breaking coupling $\kappa$ is explicit and the scaling $1/M_N$ is reminiscent of the familiar tree level Seesaw mechanism.
To get a feeling for the involved scales let us estimate the neutrino mass in the single generation limit
\begin{equation}\label{eq:numass}
    m_\nu \simeq \SI{0.1}{\electronvolt}\cdot \left(\frac{Y_{LN}}{0.1} \right)\cdot \left(\frac{Y_{
    NR}}{0.1} \right)\cdot \left(\frac{\kappa}{\SI{1}{\tera\electronvolt}} \right) \cdot \left(\frac{10^{11}\;\text{GeV}}{M_N} \right) \cdot \left(\frac{\text{Log}\left(\frac{M_N^{2}}{m_0^2}\right)-1}{\mathcal{O}(10)}\right),
\end{equation}
where in the above we used $m_0= \mathcal{O}\left(\SI{1}{\tera\electronvolt}\right)$.
Constraints on this scenario from lepton flavour violation and collider searches  can be found in \cite{Guo:2020qin}. Note that since we will investigate a different implementation of Dark matter compared to the usual Scotogenic idea, we can push the masses of the scalars and $N$ to values (far) above the electroweak scale, avoiding all laboratory constraints.
 
\subsection{Extension for radiative DM mass}\label{sec:ext}

\noindent We proceed by introducing four Weyl fermions which are chiral under $\text{U}(1)_\text{B-L}$.
Usually one charges  three right handed neutrinos with $Q_\text{B-L}=1$ so they form a vector-like pair with the $\nu_L$ from the leptonic doublet ( the $e_L$ form a vector-like pair with $e_R$). However there are other anomaly free choices such as two right handed neutrinos with  $Q_\text{B-L}=-4$ accompanied by another one with $Q_\text{B-L}=5$.
The idea of having chiral charges was originally put forth in \cite{Montero:2007cd}  and applied to dark matter in \cite{Sanchez-Vega:2014rka,Ma:2014qra,Sanchez-Vega:2015qva,Ma:2015mjd,Patra:2016ofq}. Here we propose a new realization of this idea:
Two Weyl fermions will be right handed and of equal charge $Q_1$ in order to form two massive Dirac fermions with $\nu_L$. Therefore our model predicts that the third SM neutrino remains exactly massless.
The remaining two fermions will be the right handed $\chi_3$ and the left handed $\chi_4$, which combine to form a Dirac fermion, that will be identified with the dark matter candidate. Since we have a gauge symmetry in mind, we need to find an anomaly free set of charges. As we only consider SM singlets with chiral charges there are only two  conditions for cancelling the gravitational and $\text{U}(1)_\text{B-L}^3$ anomalies from the Standard Model:
\begin{eqnarray}
    \sum_\text{dark sector} Q_\text{B-L} &= -2 Q_1 -Q_3 + Q_4 \overset{!}{=}3 \\
     \sum_\text{dark sector} Q_\text{B-L}^3 &= -2 Q_1^3 -Q_3^3 + Q_4^3 \overset{!}{=}3 
\end{eqnarray}
Here the signs reflect the fact that only $\chi_4$ is left handed. The system of equations is under-determined and has infinitely many solutions. In order for the same one-loop topology and soft breaking to generate the dark matter mass term $\overline{\chi_4}\chi_3\equiv \overline{\chi_L}\chi_R$ we impose the additional condition 
\begin{equation}
    \left|1+Q_1\right| = \left|Q_3-Q_4\right|.
\end{equation}
Without the absolute value we find no solutions.
For $1+Q_1=-(Q_3-Q_4)$ we find two possible solutions with irrational charges
\begin{equation}\label{eq:irr1}
    Q_1 = -2, \quad Q_3 =\frac{1-\sqrt{17}}{2}, \quad Q_4 =-\frac{1+\sqrt{17}}{2},
\end{equation}
and
\begin{equation}\label{eq:irr2}
    Q_1 = -2, \quad Q_3 =\frac{1+\sqrt{17}}{2}, \quad Q_4 =-\frac{1-\sqrt{17}}{2}.
\end{equation}
One can see that both sets of solutions are related by exchanging $Q_3\leftrightarrow -Q_4$.
The only solution possible for $3$ copies of  $\nu_R$ with the same charge $Q_1$ would be $Q_1=-1$ and $Q_3=Q_4$, which would allow for a term $\overline{L}\epsilon H^\dagger \nu_R$ at tree level and hence will not be investigated further. This is why our model predicts only two massive SM neutrinos. Note that formal quantum-gravitational conjectures \cite{Banks:2010zn} seem to exclude abelian gauge theories with irrational charges in curved space-time. We do not consider this line of reasoning further for our purely phenomenological study. \newline
Let us emphasize that for this particle content we need a soft breaking by $|1+Q_1|=1$ unit. However in that case any of the previously mentioned unwanted mass terms could arise at the loop level via some number of insertions of the trilinear term. Furthermore since we break the gauge symmetry by only one unit, there will be no residual $\mathcal{Z}_N$ symmetry that also stabilizes the dark matter. To remedy both shortcoming we resort to imposing  an ad-hoc $\mathcal{Z}_5$ symmetry as well. The choice of an odd $N$ was motivated by the need to forbid bilinear terms. All the charges and representations to realize the original Dirac Scotogenic model \cite{Gu:2007ug,Farzan:2012sa} with our exotic choice of $\text{U}(1)_\text{B-L}$ charges can be found in the table \ref{tab:charges-reps}.
\newline
Let us focus on the dark matter mass now: Motivated by Zee's model \cite{Zee:1980ai} for neutrino masses we consider the second topology depicted in figure \ref{fig:mass-nu-DM}. We add a pair of vector-like doublets $(D_{L},D_R)$ with $Y=-1/2$ together with another pair of vector-like singlets $(S_L,S_R)$ with $Y=0$:
\begin{equation}\label{eq:BSMYuk2}
    \mathcal{L} \supset -Y_{\chi D}\;\overline{\chi_L} \eta \epsilon D_R - Y_{S\chi}\;\overline{S_L}\sigma^* \chi_R - M_D \overline{D_L} D_R - M_S \overline{S_L}S_R
\end{equation}
Here we coupled the fermion $\chi$ to $\eta, \sigma^*$ instead of $\eta^\dagger,\sigma$, which was the case for the active neutrinos, because we need a soft breaking by plus one unit of B-L, whereas the active neutrinos needed a breaking by $-1$.
Since both components of $\chi$ are $\text{SU}(2)_\text{L}$ singlets unlike for the SM leptons, we do not only need a chirality flip on the internal fermion line but an insertion of the Higgs doublet as well:
\begin{equation}
    \mathcal{L} \supset -Y_{DS}\; \overline{D_L} \epsilon H^\dagger S_R - Y_{SD}\; \overline{S_L} H \epsilon D_R
\end{equation}
These couplings are the reason why all $D,S$ have the common B-L charge $1+Q_3$ see table \ref{tab:charges-reps}. B-L forbids all Majorana masses $\overline{S_L^c} S_L$, $\overline{S_R^c} S_R$, $\overline{\chi_L^c}\chi_L, \overline{\chi_R^c}\chi_R$ as they need a breaking by $2(1+Q_3),2Q_3, 2Q_4$ units. Since $Q_{3,4}$ are irrational numbers, no loop graph with an arbitrary number of soft symmetry breaking insertions by one unit can ever accidentally produce these terms. Hence we will leave $\chi_{L,R}$ uncharged under the $\mathcal{Z}_5$. This automatically forbids any mass mixing between $\chi$ and the $\nu_{L,R}$ as well as the kinematically allowed radiative decay $\chi\rightarrow \nu \gamma$ or the three-body decay $\chi\rightarrow \nu \overline{\nu}\nu$. Consequently the DM candidate will be absolutely stable. All other mass terms of the schematic form $L D$, $D H^\dagger e_R$, $S N$, $S \nu_R$, $S \chi$, $N \chi$ are each forbidden by at least one of the symmetries or both.
\newline
The dark matter mass term from figure \ref{fig:mass-nu-DM} depends on the mass mixing in the scalar sector as well as on the mixing between the $D$ and $S$. Their mass matrix reads
\begin{equation}
    \left(\overline{S_L},\overline{D_L^0}\right)\cdot \begin{pmatrix} M_S & -\frac{Y_{SD}v_H}{\sqrt{2}}\\
    \frac{Y_{DS}v_H}{\sqrt{2}}& M_D\end{pmatrix}\cdot \begin{pmatrix}S_R\\D_R^0    \end{pmatrix}
\end{equation}
and we find the following eigenvalues
\begin{equation}
    M_{1,2} = \frac{1}{2}\left(M_D+M_S\mp \sqrt{\left(M_D-M_S\right)^2-2v_H^2 Y_{DS} Y_{SD}}\right).
\end{equation}
The diagonalization simplifies in the limit $Y_{SD} =-Y_{DS}$ and we arrive at
\begin{equation}
    \begin{pmatrix} S_L\\D_L^0\end{pmatrix} = \begin{pmatrix} \cos(\beta)& \sin(\beta) \\-\sin(\beta) & \cos(\beta) \end{pmatrix}\begin{pmatrix}\left(F_1\right)_L\\ \left(F_2\right)_L\end{pmatrix},
    \;
    \begin{pmatrix} S_R\\D_R^0\end{pmatrix} = \begin{pmatrix} \cos(\beta)& \sin(\beta) \\-\sin(\beta) & \cos(\beta) \end{pmatrix}\begin{pmatrix}\left(F_1\right)_R\\ \left(F_2\right)_R\end{pmatrix},
\end{equation}
with a mixing angle
\begin{equation}\label{eq:ferm-mixing}
    \sin(2\beta) = \frac{\sqrt{2}Y_{DS} v_H}{2 \Delta M_F}, \quad \text{where} \quad \Delta M_F \equiv \frac{M_2-M_1}{2}.
\end{equation}
The dark matter mass arises due to eight loop diagrams in the mass basis. Since $S_i$ and $A_i$ are mass degenerate there will be only four distinct kinds of diagrams. 
For a fixed intermediate $F_j$ there are two diagrams depending on $S_1(A_1)$ and $S_2(A_2)$ again with a relative sign. Consequently all divergences will cancel in the sum and the resulting DM mass is finite.
For a fixed intermediate $S_i(A_i)$ there are two possible diagrams involving $F_1$ and $F_2$, both with a relative sign due to the fermionic mass mixing. This explains the structure of the expression for the DM mass:
\begin{equation}
    m_\text{DM} = -\frac{Y_{\chi D}Y_{S\chi}}{128 \pi^2}\sin(2\alpha) \sin(2\beta) \sum_{j=1}^2 C_j \left[ \frac{m_2^2\; \text{Log}\left(\frac{m_2^2}{M_{j}^2}\right)}{m_2^2-M_j^2}- \frac{m_1^2\; \text{Log}\left(\frac{m_1^2}{M_j^2}\right)}{m_1^2-M_j^2} \right] 
\end{equation}
with $ C_2=-C_1=1$. By working in the radiative seesaw limit 
\begin{equation}\label{eq:limit2}
    M_F \equiv \frac{M_2+M_1}{2} \gg m_0, \; \Delta M_F 
\end{equation}
and invoking the definition of the mixing angles \eqref{eq:sc-mixing} and \eqref{eq:ferm-mixing} we finally obtain
\begin{equation}\label{eq:DMmass}
    m_\text{DM} = \frac{Y_{\chi D} Y_{DS} Y_{S\chi}}{\sqrt{2}\;32\pi^2} \frac{\kappa v_H^2}{M_F^2} \left(\text{Log}\left(\frac{M_F^2}{m_0^2}\right)-3\right).
\end{equation}
Note that since we generate the dark matter mass via a dimension five operator compared to the active neutrinos (see figure \ref{fig:mass-nu-DM}), whose mass is an effective dimension four operator, there is another inverse power of the heavy suppression scale $M_F$ when compared to \eqref{eq:act-mass}. Because we want our dark matter to be heavier than the neutrinos we therefore need $M_N \gg M_F$, which can be seen from the following estimate 
\begin{equation}\label{eq:mDM}
    m_\text{DM} \simeq  \SI{4}{\kilo\electronvolt}\cdot \left(\frac{Y_{\chi D}}{0.1} \right)\cdot \left(\frac{Y_{DS}}{0.1} \right)\cdot \left(\frac{Y_{S \chi}}{0.1} \right) \cdot 
    \left(\frac{\kappa}{\SI{1}{\tera\electronvolt}} \right) \cdot \left(\frac{\SI{30}{\tera\electronvolt}}{M_F} \right)^2 \cdot \left(\frac{\text{Log}\left(\frac{M_F^{2}}{m_0^2}\right)-3}{3}\right).
\end{equation}
In the above we used $m_0= \mathcal{O}\left(\SI{1}{\tera\electronvolt}\right)$. Unlike the $N$ which are much heavier the $F$ and electrically charged components of $D$ could be potentially be produced at future colliders and have a direct coupling to the SM like Higgs. 

\subsection{UV completion}\label{sec:UV}
\noindent In order to gauge the $\text{U}(1)_\text{B-L}$ symmetry and to explain the origin of the dimensionful coupling $\kappa$ in the  trilinear term \eqref{eq:tril} we introduce a second SM singlet scalar $\phi$ with the charge $Q_\phi = -1 -Q_1= 1$ without any couplings to the fermion spectrum:
\begin{eqnarray}
    \mathcal{L}_\phi\supset &-& \mu_\phi^2 \left|\phi\right|^2  - \lambda_\text{IV} \left(\eta^\dagger H \sigma \phi^* +\; \text{h.c.}\right)\\
    &-&\lambda_\phi \left|\phi\right|^4- \lambda_{H\phi} \left(H^\dagger H\right)  \left|\phi\right|^2   
     - \lambda_{\eta \phi }  \left(\eta^\dagger \eta\right)  \left|\phi\right|^2   -
    \lambda_{\sigma \phi } \left|\sigma\right|^2 \left|\phi\right|^2.
\end{eqnarray}
We parameterize the new scalar as 
\begin{equation}
    \phi = \frac{\phi_R^0 + v_\phi +  i \phi_I^0}{\sqrt{2}},
\end{equation}
which allows us to identify $\kappa= \lambda_\text{IV} v_\phi \equiv \lambda_\text{IV} v_\text{B-L}$. We do not depict an insertion of this vev in figure \ref{fig:mass-nu-DM}, because the neutrino and DM mass generation only requires a non-zero value of $\kappa$ irrespective of its origin in the UV.
The $\phi_I^0$ is the would-be-Goldstone-Boson that gets absorbed to become the longitudinal component of the massive $\text{U}(1)_\text{B-L}$ gauge boson that we call $Z'$ whose mass reads
\begin{equation}\label{eq:vevbound}
    m_{Z'} = g_\text{B-L} v_\text{B-L},
\end{equation}
because $\phi$ is the only field with B-L charge that receives a vev.
Direct searches at LEP place the following bound \cite{PhysRevD.70.093009,Heeck:2014zfa} on the mass of a new gauge boson
\begin{equation}\label{eq:vev-lab}
    v_\text{B-L} = \frac{m_{Z'}}{g_\text{B-L}} > \SI{6.9}{\tera\electronvolt}\quad @ \; 95\%\; \text{C.L.}
\end{equation}
that couples to the  conventional B-L charges of the SM fermions. Searches at the LHC exclude $Z'$s below $0.2-\SI{3.5}{\tera\electronvolt}$ \cite{ATLAS:2014pcp}. Since no scalar field that receives a vev is charged under both weak isospin/hypercharge or  B-L there is no mass mixing between the Z and Z' bosons. However there can be gauge kinetic mixing \cite{Holdom:1985ag}, for instance generated at the loop level by self-energy graphs containing the $(D_L,D_R)$ or $\eta$ fields, which are charged under both abelian symmetries and weak isospin.\newline
The additional scalar interactions contribute to the masses of the $\eta^0$ and $\sigma^0$ bosons by shifting the relations in \eqref{eq:mass} to
\begin{eqnarray}
    \tilde{m}_\eta^2 \rightarrow   m_\eta^2 + \left(\lambda_{H\eta\;1}+\lambda_{H\eta\;2}\right)v_H^2 + \lambda_{\eta \phi }   v_\text{B-L}^2, \; \text{and} \;  \tilde{m}_\sigma^2 \rightarrow  m_\sigma^2 + \lambda_{H\sigma}v_H^2 + \lambda_{\sigma \phi } v_\text{B-L}^2.
\end{eqnarray}
Additionally the mixed quartic between $H$ and $\phi$ leads to mass mixing between them: First we minimize the potential in each direction and find expressions to eliminate the parameters $\mu_H^2, \mu_\phi^2 <0$. We find that the minimum in each direction can be obtained for
\begin{equation}
    \frac{\mu_H^2}{2} = - 2 \lambda_H v_H^2 - \lambda_{H\phi} v_\text{B-L}^2, \quad \text{and} \quad  \frac{\mu_\phi^2}{2} = - 2 \lambda_\phi  v_\text{B-L}^2 - \lambda_{H\phi}v_H^2 
\end{equation}
and we arrive at
\begin{equation}
  \left(h_R,\phi_R^0\right) \cdot \begin{pmatrix}2 \lambda_H v_H^2  & \frac{\lambda_{H\phi}}{2} v_H  v_\text{B-L} \\ \frac{\lambda_{H\phi}}{2} v_H  v_\text{B-L} & 2 \lambda_\phi  v_\text{B-L}^2 \end{pmatrix}\cdot \begin{pmatrix} h_R\\ \phi_R^0\end{pmatrix},
\end{equation}
with the eigenvalues
\begin{equation}
    m_{h, \varphi}^2 = \lambda_H v_H^2 + \lambda_\phi  v_\text{B-L}^2  \mp \frac{1}{2}\sqrt{4 \lambda_H^2 v_H^4 + 4 \lambda_\phi^2  v_\text{B-L}^4 + v_H^2  v_\text{B-L}^2\left(\lambda_{H\phi}^2-8\lambda_H \lambda_\phi\right)}.
\end{equation}
In the limit $v_\text{B-L} \gg v_H$ we find at leading order 
\begin{equation}\label{eq:scal-masses}
    m_h^2 \simeq \left(2\lambda_H-\frac{\lambda_{H\phi}^2}{8\lambda_\phi}\right)v_H^2 \quad \text{and} \quad m_\varphi^2 \simeq  2\lambda_\phi v_\text{B-L}^2.
\end{equation}
The correction to the SM like Higgs mass can be understood as a tree level threshold correction to its quartic from integrating out the heavier   field \cite{Elias-Miro:2012eoi}. The mass eigenstates are determined from
\begin{equation}
    \begin{pmatrix}  h_R \\ \phi_R^0\end{pmatrix} = \begin{pmatrix} \cos(\gamma)& \sin(\gamma) \\-\sin(\gamma) & \cos(\gamma) \end{pmatrix}\begin{pmatrix}h \\ \varphi \end{pmatrix}
\end{equation}
with
\begin{equation}
    \sin(2\gamma) = \frac{\lambda_{H\phi}v_H v_\text{B-L}}{2\Delta m_h^2}, \quad \text{where} \quad \Delta m_h^2 \equiv \frac{m_\varphi^2-m_h^2}{2}
\end{equation}
and at leading order in $v_H/v_\text{B-L}$ this reduces to 
\begin{equation}
    \sin(2\gamma) \simeq  \frac{\lambda_{H\phi}}{2\lambda_\phi} \cdot \frac{v_H}{v_\text{B-L}}.
\end{equation}
In the present study we will neglect this mixing completely. It is important to note that the discrete $\mathcal{Z}_5$ symmetry we imposed will most likely be broken by quantum gravitational effects \cite{COLEMAN1988643,GIDDINGS1988854,GILBERT1989159}, which is why we assume it is e.g. a residual symmetry arising from the spontaneous symmetry breaking of a gauge symmetry \cite{doi:10.1063/1.1665530}. This larger symmetry could also connect our choice of $\text{U}(1)_\text{B-L}$ with the rest of the SM gauge group, e.g. by unifying it with QCD into the Pati-Salam hypercolor $\text{SU}(4)_\text{c}$ \cite{PhysRevD.10.275}.  Vector-like fermions such as our singlets $(N_L,N_R)$ and $(S_L,S_R)$, doublets $(D_L,D_R)$ as well as exotic vector-like down-type quarks arise in $\text{E}_6$-GUTs \cite{GURSEY1976177,SHAFI1978301}. This could provide an interesting route for further completing our model in the UV as the Pati-Salam model can be embedded in $\text{SO}(10)$ which is a subgroup of $\text{E}_6$.

\section{Dark Matter}
\noindent
As previously mentioned our DM candidate does not mix with the active neutrinos. Hence the usually considered possibility of creating keV-scale neutrino DM via active-to-sterile oscillations \cite{Dodelson:1993je}, that can be enhanced in the presence of a chemical potential for neutrinos \cite{Shi:1998km}, are ruled out and we have to look into other avenues to produce DM.
In the following we will briefly explain why we do not consider thermal production and focus on non-thermal scenarios. To study non-thermal production of DM we assume that the reheating temperature $T_\text{RH}$ of the universe is below both the masses of the particles in the loops of figure \ref{fig:mass-nu-DM}   and the mass of the B-L gauge boson $Z'$
\begin{equation}
    M_N\gg M_F\gg m_0 \gg T_\text{RH} \quad \text{and} \quad m_{Z'}\gg T_\text{RH}.
\end{equation}
This ensures that none of the new, potentially stable neutral particles, which are good thermal dark matter candidates, are present in the plasma. We can thus limit ourselves to the SM degrees of freedom augmented by the two $\nu_R$ and the light DM.

\subsection{Lyman bound for FIMPs}\label{sec:lyman}

\noindent The Lyman-$\alpha$ forest consists of absorption lines in the spectra of quasars due to neutral hydrogen in the intergalactic medium. It provides a window into the matter power spectrum, which  contains information on the Dark matter's free-streaming scale from the time of structure formation. One can use the existing data on the Lyman-$\alpha$ forest to set bounds on dark matter models affecting small scale structures such as the thermally produced warm DM (WDM). Numerically challenging simulations for WDM have been performed and lead to a lower limit of $m_\text{WDM}^{\text{Ly-}\alpha}= \SI{5.3}{\kilo\electronvolt}$ at 95\% confidence level (CL) \cite{Viel:2013fqw,Palanque-Delabrouille:2019iyz}. Reference  \cite{Garzilli:2019qki} argued that the aforementioned bound is too strong when taking into account systematics such as assumptions about the thermal history and instead they find $m_\text{WDM}^{\text{Ly-}\alpha}= \SI{1.9}{\kilo\electronvolt}$ at 95\% CL.
In order to avoid such time consuming simulations for other DM production modes a bound mapping formalism has been devised in \cite{Bae:2017dpt,Murgia:2017lwo,Heeck:2017xbu,Boulebnane:2017fxw,Baldes:2020nuv,Ballesteros:2020adh,DEramo:2020gpr} and a recent reevaluation \cite{Decant:2021mhj} found that the previously mentioned mass range $m_\text{WDM} \gtrsim (1.9-5.3)\;\text{keV}$ translates into a bound on the FIMP mass of $m_\text{FIMP} \gtrsim (4-16)\;\text{keV}$.

\subsection{Out of equilibrium Higgs decays}\label{sec:higgs}
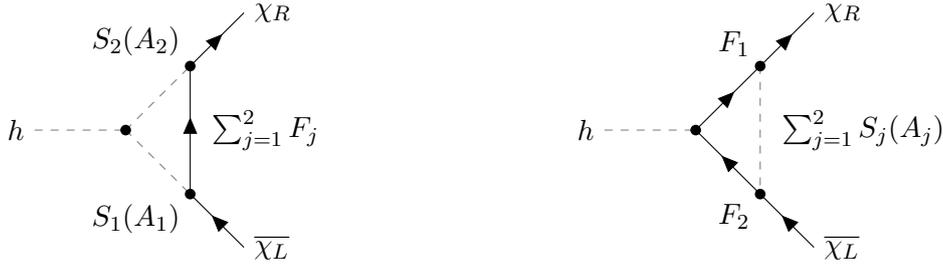
\begin{figure}[t!]
 \centering
  \tikzset{
  blackline/.style={thin, draw=black, postaction={decorate},
    decoration={markings, mark=at position 0.6 with {\arrow[black]{triangle 45}}}},
    blueline/.style={thin, draw=blue, postaction={decorate},
    decoration={markings, mark=at position 0.6 with {\arrow[blue]{triangle 45}}}},
    redline/.style={thin, draw=red, postaction={decorate},
    decoration={markings, mark=at position 0.6 with {\arrow[red]{triangle 45}}}},
    greenline/.style={thin, draw=green, postaction={decorate},
    decoration={markings, mark=at position 0.6 with {\arrow[green]{triangle 45}}}},    
    graydashed/.style={dashed, draw=gray, postaction={decorate},
    decoration={markings}},
   yellowdashed/.style={dashed, draw=orange, postaction={decorate},
    decoration={markings}},
    photon/.style={decorate, draw=red,
    decoration={coil,amplitude=12pt, aspect=0}},
    reddashed/.style={thick, dashed, draw=red, postaction={decorate},
    decoration={markings}},
    photon/.style={decorate, draw=red,
    decoration={coil,amplitude=12pt, aspect=0}},
  gluon/.style={dashed, decorate, draw=black,
    decoration={coil, segment length=5pt, amplitude=8pt}}
  line/.style={thick, draw=black, postaction={decorate},
    decoration={markings}}
}

\NewDocumentCommand\semiloop{O{black}mmmO{}O{above}}
{%
\draw[#1] let \p1 = ($(#3)-(#2)$) in (#3) arc (#4:({#4+180}):({0.5*veclen(\x1,\y1)})node[midway, #6] {#5};)
}

\begin{tikzpicture}[node distance=1cm and 1cm]

\coordinate[label = left: $h$ ] (start1);
\coordinate[right=1.2cm of start1] (vertex1);
\coordinate[above right =1.2cm of vertex1,label=above left: $S_2(A_2)$] (vertex2);
\coordinate[below right =1.2cm of vertex1,label=below left: $S_1(A_1)$] (vertex3);
\coordinate[above right =1cm of vertex2,label=right: $\chi_R$ ] (end2);
\coordinate[below right =1cm of vertex3,label=right: $\overline{\chi_L}$ ] (end3);
\coordinate[right =1cm of vertex1,label=right: $\sum_{j=1}^2 F_j$] (label1);

\draw[graydashed] (start1)   -- (vertex1);
\draw[graydashed] (vertex1)   -- (vertex2);
\draw[graydashed] (vertex1)   -- (vertex3);
\draw[blackline] (end3) -- (vertex3);
\draw[blackline] (vertex3) -- (vertex2);
\draw[blackline] (vertex2) -- (end2);

\fill (vertex1) circle (2pt);
\fill (vertex2) circle (2pt);
\fill (vertex3) circle (2pt);

\coordinate[right=7.5cm of start1,label = left: $h$ ] (start21);
\coordinate[right=1.2cm of start21] (vertex21);
\coordinate[above right =1.2cm of vertex21,label=above left: $F_1$] (vertex22);
\coordinate[below right =1.2cm of vertex21,label=below left: $F_2$] (vertex23);
\coordinate[above right =1cm of vertex22,label=right: $\chi_R$ ] (end22);
\coordinate[below right =1cm of vertex23,label=right: $\overline{\chi_L}$ ] (end23);
\coordinate[right =1cm of vertex21,label=right: $\sum_{j=1}^2 S_j (A_j)$] (label21);

\draw[graydashed] (start21)   -- (vertex21);
\draw[graydashed] (vertex22) -- (vertex23);
\draw[blackline] (vertex22) -- (end22);
\draw[blackline] (vertex21) -- (vertex22);
\draw[blackline] (vertex23) -- (vertex21);
\draw[blackline] (end23) -- (vertex23);

\fill (vertex21) circle (2pt);
\fill (vertex22) circle (2pt);
\fill (vertex23) circle (2pt);

\end{tikzpicture}
  \caption{Leading order diagrams for the decay $h\rightarrow \overline{\chi_L}\chi_R$ in the mass basis. See the main text for more details. }
  \label{fig:decays}
\end{figure}

\noindent
In the following we focus on the decay $h\rightarrow \overline{\chi_L}\chi_R, \overline{\chi_R}\chi_L $, which is obtained from the second diagram in \ref{fig:mass-nu-DM} by replacing one of the Higgs vevs with the radial excitation $h$, which leads to the two diagrams depicted in figure \ref{fig:decays}. By replacing both Higgs vevs one can compute the scattering process $h h \rightarrow \overline{\chi_L}\chi_R, \overline{\chi_R}\chi_L $, however for our first estimate we will limit ourselves to the decays. We only consider the trilinear coupling from \eqref{eq:tril} and neglect all the decay modes to the same chiralities  ($LL$ or $RR$) which occur via the quartic couplings $\lambda_{H\eta 1,2},\lambda_{H\sigma}$, from DM mass insertions on the external lines \cite{Ma:2021eko}, or from mass mixing in the heavy scalar or fermion sector  to focus on the parameters for the DM mass. This is also why we will work to the lowest order in the mixing angles $\sin(\alpha)$ and $\sin(\beta)$ because in the mass basis there are 32 diagrams contributing and both neutrino masses were independent of the aforementioned angles in the radiative seesaw limit.  We neglect the mixing $\sin(\gamma)$ between $h$ and $\varphi$. In this approximation with $\cos(\alpha)\simeq \cos(\beta)\simeq 1$ and $S_1\simeq \eta_R^0, \; S_2\simeq \sigma_R^0$, $F_1\simeq S,\; F_2 \simeq D^0 $ there are only four diagrams contributing. 
By dropping  the final state DM mass we find
\begin{equation}
    \Gamma(h\rightarrow \overline{\chi}\chi) =\Gamma(h\rightarrow \overline{\chi_L}\chi_R) + \Gamma(h\rightarrow  \overline{\chi_R}\chi_L)= \frac{m_h}{8\pi} \left|f_\nu\right|^2.
\end{equation}
The first set of graphs depicted on the left side of figure \ref{fig:decays} is obtained by replacing the upper vev in \ref{fig:mass-nu-DM}  with $h$ and only depends on $\sin(\beta)$.
The amplitude is finite because it comes from a difference of terms due to the relative sign between the  $F_{1,2}$ contributions. The corresponding effective Yukawa coupling is found to be at leading order in $\sin(\beta)$ from \eqref{eq:sc-mixing} and by making use of \eqref{eq:limit2}
\begin{equation}\label{eq:falpha}
    f_\beta = \frac{Y_{\chi D}Y_{DS} Y_{S\chi}}{\sqrt{2}\;2} \frac{\kappa v_H}{M_F^2}\left(\text{Log}\left(\frac{M_F^{2}}{m_0^2}\right)-2\right).
\end{equation}
Similarly the second diagram on the right side of figure \ref{fig:decays} is obtained by replacing the lower vev in \ref{fig:mass-nu-DM} with $h$. It is proportional to $\sin(\alpha)$ from \eqref{eq:ferm-mixing} and finite because here the difference arises due to the relative sign of the $S_{1,2}(A_{1,2})$ contributions. The effective coupling is
\begin{equation}
    f_\alpha= \frac{Y_{\chi D}Y_{DS} Y_{S\chi}}{\sqrt{2}\;2} \frac{\kappa v_H}{M_F^2}\left(\text{Log}\left(\frac{M_F^{2}}{m_0^2}\right)-3\right).
\end{equation}
In both expressions we neglected the Higgs mass.
The sum of both contributions can be re-expressed by comparison with \eqref{eq:DMmass} as
\begin{equation}
    f_\nu = f_\alpha+ f_\beta= 2 \frac{m_\text{DM}}{v_H}\frac{\text{Log}\left(\frac{M_F^{2}}{m_0^2}\right)-\frac{5}{2}}{\text{Log}\left(\frac{M_F^{2}}{m_0^2}\right)-3} \rightarrow  2 \frac{m_\text{DM}}{v_H}\quad \text{for} \quad M_F \gg m_0 \gg m_h,
\end{equation}
which agrees with the EFT expectation that after EWSB the diagram on the right in figure \ref{fig:mass-nu-DM} can be represented by an effective Weinberg-type operator \cite{WeinbergOperator} at energy scales below all the mediator masses
\begin{equation}\label{eq:EFT}
    \mathcal{L}_\text{EFT} = 2\frac{m_\text{DM}}{v_H^2}  \overline{\chi}\chi \left(H^\dagger H\right).
\end{equation}
The remainder of this section discusses how to produce DM from this effective operator and can be applied to other models that generate this operator as well. For the decay width we find after neglecting the phase space suppression
\begin{equation}\label{eq:higgsdecay}
    \Gamma\left(h\rightarrow \overline{\chi}\chi\right) = \frac{m_h}{2\pi} \left(\frac{m_\text{DM}}{v_H}\right)^2
\end{equation}
and we emphasize that the only free parameter is the DM mass. The experimental limit on the branching ratio (BR) from searches for  invisible Higgs decays beyond the SM is between 19\% (CMS) and  26\% (ATLAS) which translates to approximately  $\Gamma\left(h\rightarrow \text{Inv.}\right) < \SI{1.3}{\mega\electronvolt}$ \cite{ATLAS:2018bnv,Cepeda:2019klc,ATLAS-CONF-2020-008}, implying an upper bound on the DM mass of roughly
\begin{equation}\label{eq:DMBound1}
    m_\text{DM} \lesssim \SI{2}{\giga\electronvolt},
\end{equation}
which justifies neglecting the phase space suppression.
Multiple proposed next generation collider experiments are expected to tighten the bound on the invisible BR by up to two orders of magnitude to $\text{BR}\left(h\rightarrow \text{Inv.}\right) = 0.22\%$ (FCC-ee) \cite{deBlas:2019rxi}, 0.24\% (CEPC) \cite{Tan:2020ufz} and 0.26\% (ILC) \cite{Ishikawa:2019uda}. The corresponding bound on the DM mass would read approximately
\begin{equation}
    m_\text{DM} \lesssim \SI{170}{\mega\electronvolt}.
\end{equation}
This bound is only one order of magnitude stronger than \eqref{eq:DMBound1} due to quadratic dependence of the branching ratio on the DM mass. The invisible Higgs decays lead to the strongest terrestrial bound on the DM mass, however as we will see  avoiding cosmological over-production of DM from Higgs mediated scatterings firmly requires the DM mass to be below the MeV-scale see \eqref{eq.bound}.\\
\\
\noindent In the following we will limit ourselves to the era of radiation domination and make extensive use of the Hubble rate and the entropy density
\begin{equation}
    H(T) \simeq 1.66 \sqrt{g_{*\rho}(T)} \frac{T^2}{M_\text{Pl.}}, \quad s(T) = g_{*S}(T) \frac{2\pi^2}{45}T^3,
\end{equation}
where $g_{*\rho}$ and $g_{*S}$ are the effective number of degrees of freedom in energy and entropy respectively.
Before we deal with non-thermal DM production let us take a look the thermal case first:
The decay \eqref{eq:higgsdecay} will be in thermal equilibrium at $T=m_h$ provided that $m_\text{DM}\gtrsim \SI{4.5}{\kilo\electronvolt}$ (we will show this later in \eqref{eq:dm-eq}). 
Since during radiation domination we have $\Gamma / H \sim T^{-2}$ for decays at temperatures below the mass of the decaying particle, a decay never falls out of thermal equilibrium. Consequently we need to know when the inverse decay freezes-out in order to find the decoupling temperature of $\chi$. The corresponding rate reads at $T\ll m_h$  \cite{Kolb:1979qa, Buchmuller:2005eh}
\begin{eqnarray}
    \Gamma_\text{ID}  &=&\frac{1}{3 \zeta(3)}\sqrt{\frac{\pi}{2}} \left(\frac{m_h}{T}\right)^\frac{3}{2} \cdot  e^{-\frac{m_h}{T}} \cdot  \Gamma\left(h\rightarrow \overline{\chi}\chi\right)
\end{eqnarray}
and the phase suppression is encoded in the Boltzmann factor. Numerically we find that this interaction freezes out at $T_\text{FO}\gtrsim \SI{3}{\giga\electronvolt}$ for $m_\text{DM}\lesssim \SI{2}{\giga\electronvolt}$. Of course there is also a scattering process $hh\rightarrow \overline{\chi}\chi$, but since this requires two on shell Higgses the rate density will be double Boltzmann-suppressed below $m_h$ typically leading to an earlier freeze-out than the inverse decays. Since $\chi$ is relativistic at decoupling it would be a warm DM candidate, however it has long been known, that such a DM candidate would overclose the universe \cite{Bezrukov:2009th}
\begin{equation}
    \Omega^\text{warm}_\text{DM} h^2  \simeq \mathcal{O}\left(10^6\right) \cdot \left(\frac{84}{g_*\left(T_\text{FO}\simeq \SI{3}{\giga\electronvolt}\right) }\right) \cdot \left(\frac{m_\text{DM}}{\SI{1}{\giga\electronvolt}}\right),
\end{equation}
if there is no release of entropy that dilutes the relic density to the observed value. Realizing the warm DM scenario requires additional degress of freedom in the plasma like long-lived particles that decoupled while relativistic whose decays generate the necessary entropy dilution \cite{PhysRevD.31.681}. For the sake of minimality we do not consider this idea further and focus on out-of-equilibrium-processes involving only SM states that are connected to the DM via the previously introduced BSM Yukawa and gauge interactions.
\newline
Next we investigate out of equilibrium Higgs decays. We use the notation of \cite{Giudice:2003jh} to write down the Boltzmann equations for the DM production where $Y_\text{DM} \equiv \frac{n_\text{DM}}{s}$, with $s$ being the entropy density and $z=\frac{m_h}{T}$
\begin{equation}
    z H s \frac{\text{d}Y_\chi}{\text{d}z} = \gamma_{h\rightarrow \overline{\chi}\chi} \frac{Y_h}{Y_h^\text{e.q.}}- \gamma_{ \overline{\chi}\chi\rightarrow h}\frac{Y_\chi}{Y_\chi^\text{e.q.}}\frac{Y_{\overline{\chi}}}{Y_{\overline{\chi}}^\text{e.q.}},
\end{equation}
where we assumed that entropy is conserved. Note that away from thermal equilibrium the temperatures of the SM and DM baths are different so that $\gamma_{h\rightarrow \overline{\chi}\chi}$ depends on $T_\text{SM}$ and $\gamma_{ \overline{\chi}\chi\rightarrow h}$ is a function of $T_\text{DM}$.
The freeze-in regime  \cite{Hall:2009bx} is defined by the condition $Y_\chi\ll Y_\chi^\text{e.q.}$ and the same for $\overline{\chi}$. If we use the fact that the SM Higgs is kept in thermal equilibrium $Y_h\simeq Y_h^\text{e.q.}$ until $T_\text{FO}\simeq \frac{m_h}{25}\simeq \SI{5}{\giga\electronvolt}$ we obtain 
\begin{equation}
    z H s \frac{\text{d}Y_\chi}{\text{d}z} \simeq \gamma_{h\rightarrow \overline{\chi}\chi},
\end{equation}
where the thermally averaged decay width density reads \cite{Kolb:1979qa}
\begin{equation}\label{eq:Boltz1}
     \gamma_{h\rightarrow \overline{\chi}\chi} = \frac{ g_h m_h^2 T}{2\pi^2} \text{K}_1 (z)  \Gamma\left(h\rightarrow \overline{\chi}\chi\right).
\end{equation}
In this context $g_h=1$ is the spin degeneracy of the Higgs and $\text{K}_1 (z) $ denotes a modified Bessel function of the first kind. 
To ensure that we are in the freeze-in regime the decay is not allowed to thermalize which leads to the condition
\begin{equation}\label{eq:outofeq1}
    \frac{\Gamma\left(h\rightarrow \overline{\chi}\chi\right)}{H(T)}\Big|_{T=m_h}<1
\end{equation}
that can be re-expressed as a bound on the DM mass
\begin{equation}\label{eq:dm-eq}
    m_\text{DM}\lesssim \SI{4.5}{\kilo\electronvolt} \cdot \left(\frac{g_{*\rho}(m_h)}{100}\right)^\frac{1}{4},
\end{equation}
that is borderline compatible with the lower limit of the Lyman-$\alpha$ window. Under the assumption that there is no primordial abundance of DM we can integrate \eqref{eq:Boltz1} to determine the DM abundance today at $z_0$
\begin{equation}
    Y_\chi(z_0) = \mathcal{C}_h \int_{z_\text{RH}}^{z_0=\infty} \text{d} z\; \frac{z^3}{g_{*S}(z) \sqrt{g_{*\rho}(z)}} \text{K}_1 (z), \quad \text{where} \quad z_\text{RH} = \frac{m_h}{T_\text{RH}}
\end{equation}
and the factor $\mathcal{C}_h$ is a short hand for all microscopic and cosmological parameters
\begin{equation}
    \mathcal{C}_h = 1.1\times 10^{-2} \frac{m_\text{DM}^2}{ v_H^2} \frac{ M_\text{Pl}}{m_h}.
\end{equation}
We then use this to compute the energy density in dark matter by using the present day entropy density $s_0$ and the critical density $\rho_c$ \cite{Rubakov:2017xzr}
\begin{equation}\label{eq:conv}
     \Omega_\text{DM} h^2 = 2 \frac{m_\text{DM} s_0}{\rho_c}  Y_\chi(z_0) \simeq 1.1\times 10^3 \left(\frac{m_\text{DM}}{\SI{4}{\kilo\electronvolt}}\right)Y_\chi(z_0).
\end{equation}
Here the factor of two arises because our DM candidate is a Dirac fermion. For a simple analytical estimate we can neglect the temperature dependence of the relativistic number of degrees of freedom in energy $g_{* \rho}(z)$ and entropy  $g_{*S}(z)$ and replace them by their average values at the time of predominant dark matter production. This can be done because freeze-in production of DM is always sharply peaked around either $T\simeq m_h$ for the IR freeze-in \cite{Hall:2009bx,Bernal:2017kxu} or at the reheating temperature $T_\text{RH}$ for UV freeze-in \cite{Elahi:2014fsa}. First let us suppose a standard big bang cosmology that corresponds to $z_\text{RH}\rightarrow 0$ which gives the maximally possible abundance 
\begin{equation}\label{eq:max-abund}
    Y_\chi(z_0)^\text{max} \simeq \frac{4.71\;\mathcal{C}_h}{g_{*S}(m_h) \sqrt{g_{*\rho}(m_h)}}
\end{equation}
that corresponds to 
\begin{equation}
     h^2 \Omega_\text{DM} \simeq 0.12 \cdot  \left(\frac{m_\text{DM}}{\SI{1.5}{\kilo\electronvolt}}\right)^3 \cdot \left(\frac{100}{ g_{*S}(m_h)}\right) \cdot \sqrt{\frac{100}{g_{* \rho}(m_h)}},
\end{equation}
where we used the maximum possible number of relativistic degrees of freedom above the EW phase transition in the SM. One can see that the correct relic density \cite{Planck:2018vyg} is obtained for a DM mass that is in conflict with the more conservative Lyman-$\alpha$ bound that requires $m_\text{DM} > \SI{4}{\kilo\electronvolt}$. Since $ h^2 \Omega_\text{DM} \sim m_\text{DM} Y_\chi(z_0)$ we can allow for a larger DM mass by lowering the yield $ Y_\chi(z_0)$. This is most easily done by assuming $z_\text{RH}>0$ which lowers the relic abundance below \eqref{eq:max-abund}. In doing so we introduce a second free parameter in the form of $T_\text{RH}$. We find that we can decrease the abundance for $z_\text{RH}>1$, however our fix comes with two complications:
On the one hand one needs to make sure that the SM Higgs is actually thermalized after reheating. Reference \cite{Harigaya:2013vwa} found that particles charged under non-abelian gauge symmetries that are produced from inflaton decays during reheating thermalize before the end of reheating (which is not an instantaneous process) provided that the fine structure constant of the gauge interaction satisfies
\begin{equation}
    \alpha\gg \alpha_\text{Lim} \equiv \left(\frac{m_I}{M_\text{Pl}}\right)^\frac{5}{8} \cdot \left(\frac{\Gamma_I M_\text{Pl}}{m_I^3}\right)^\frac{1}{8}.
\end{equation}
In this context $m_I$ is the inflaton mass and $\Gamma_I$ is its decay width, which we can trade for an expression involving $T_\text{RH}$ (see \eqref{eq:dec-rh}). We find that 
\begin{equation}\label{eq:therm-RH}
    \alpha_\text{Lim} \simeq 2\times 10^{-9}\cdot \left(\frac{m_I}{\SI{1}{\tera\electronvolt}}\right)^\frac{1}{4}\cdot \left(\frac{T_\text{RH}}{\SI{1}{\giga\electronvolt}}\right)^\frac{1}{4} \cdot \left( \frac{g_{*\rho}\left(T_\text{RH}\right)}{76}\right)^\frac{1}{16},
\end{equation}
which is definitely satisfied for the SM Higgs coupling to $\text{SU}(2)$ gauge bosons where $\alpha_2 = \frac{g^2}{4\pi }$ with $g\simeq0.64$.
On the other hand the out of equilibrium condition \eqref{eq:outofeq1} must be re-evaluated at $T_\text{RH}<m_h$ leading to 
\begin{equation}
    m_\text{DM}\lesssim \frac{\SI{4.5}{\kilo\electronvolt}}{z_\text{RH}} \cdot \left(\frac{g_{*\rho}(z_\text{RH})}{100}\right)^\frac{1}{4}.
\end{equation}
The necessary $z_\text{RH}>1$ leads to DM masses which even  violate the  lower more conservative Lyman-$\alpha$ bound.
In other words: If we tried to satisfy the Lyman-$\alpha$ window we would obtain a thermalized population of $\chi$, which would actually be warm dark matter and this can only be made to work with additional processes that release enough entropy to dilute it. Since this channel leads to over-production of dark matter and the inclusion of $2\rightarrow 2$ scattering processes will only increase the relic abundance further, we conclude that freeze-in from the SM Higgs via a Weinberg-type operator is not a viable production mode for keV-scale DM. Furthermore in order to avoid any contribution from Higgs decays we will only consider cosmologies with $T_\text{RH}< T_\text{FO}\simeq \frac{m_h}{25}\simeq \SI{5}{\giga\electronvolt}$. Successful BBN requires a reheating temperature of at least $\SI{4}{\mega\electronvolt}$ \cite{Kawasaki:2000en,Hannestad:2004px}.

\subsection{Super WIMP contribution}
\noindent Another production channel for DM is the Super WIMP scenario \cite{Feng:2003xh} in which the DM is produced after the thermal freeze-out of the Higgs boson from its gauge and Yukawa interactions at $ T_\text{FO}\simeq \frac{m_h}{25}$. However the Higgs   has   decay modes to SM particles which are much faster than the decay to DM so the frozen out abundance of Higgses can not lead to a significant production of DM.

\subsection{Gauge Scattering}\label{sec:gauge1}
\begin{figure}[!t]
    \centering
    \begin{minipage}{0.75\textwidth}
    \includegraphics[width=\textwidth]{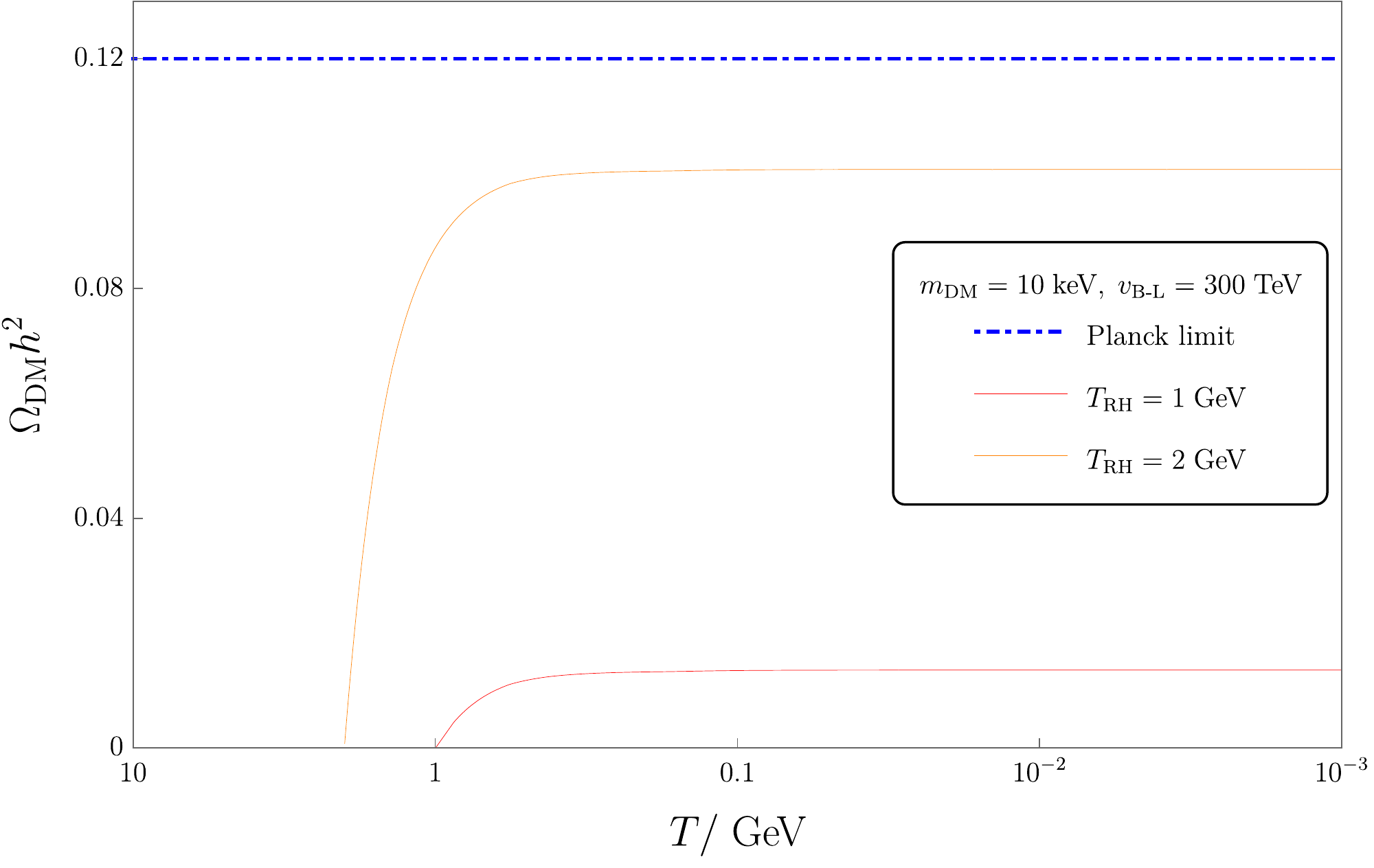}
    \caption{DM abundance  as a function of temperature for fixed $m_\text{DM}, v_\text{B-L}$ and two different $T_\text{RH}$.}
    \label{fig:YDM1}
    \end{minipage}
    
    \vspace{0.05\textwidth}
    \begin{minipage}{0.75\textwidth}
    \includegraphics[width=\textwidth]{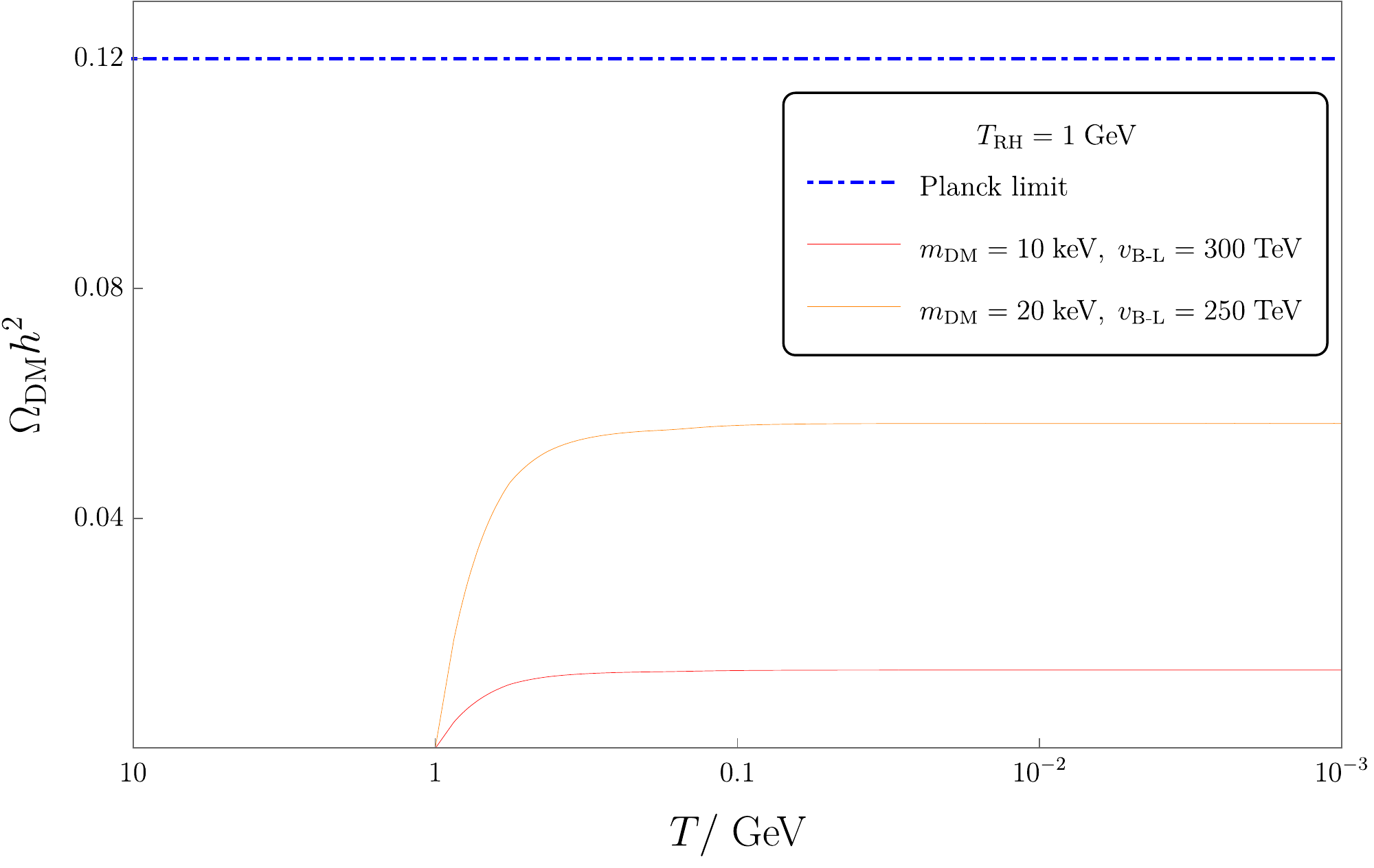}
    \caption{DM abundance  as a function of temperature for fixed $T_\text{RH}$ and two different combinations of $m_\text{DM}, v_\text{B-L}$.}
    \label{fig:YDM2}
    \end{minipage}
\end{figure}

\noindent
We can also produce DM via the new gauge interaction \cite{Khalil:2008kp,Das:2021nqj}. In the limit $s\ll m_{Z'}$ the cross section for interconverting DM and  SM fermions $f_i$ via $Z'$ exchange reads for massless fermions \cite{Barger:2003zh}
\begin{align}\label{eq:xsec}
    \sigma\left(\overline{\chi}\chi \leftrightarrow \overline{f}_i f_i\right) &\equiv  \frac{ \alpha_{\chi i}\; s}{12 \pi}\\
    &= \frac{s}{12\pi}\cdot \left(\frac{g_\text{B-L}}{m_{Z'}}\right)^4\cdot \left(Q\left(\chi_L\right)^2+Q\left(\chi_R\right)^2\right) \left(N_c\right)_i \left(Q\left(f_{i\;L}\right)^2+Q\left(f_{i\;R}\right)^2\right)\nonumber
\end{align}
where $Q$ denotes the various B-L charges and $N_c$ is a color factor which equals three for quarks and one for leptons. The above was summed and not averaged over the initial state spins.
Since $m_{Z'}= g_\text{B-L} v_\text{B-L}$ the cross section is only sensitive to $v_\text{B-L}$ in the effective operator limit. Even though the DM mass in \eqref{eq:mDM} formally depends on $\kappa =  \lambda_\text{IV} v_\text{B-L}$ we treat $m_\text{DM}$ and  $v_\text{B-L}$ as independent parameters, because a larger $v_\text{B-L}$ can always be compensated by a smaller $ \lambda_\text{IV}$ or by making the fermions running in the loop heavier.
\newline
The Boltzmann equation read for $z\equiv \frac{T_\text{RH}}{T}$
\begin{align}\label{eq:Boltz2}
    z H s \frac{\text{d}Y_\chi}{\text{d}z} &= \sum_i \gamma_{\overline{f}_i f_i\rightarrow \overline{\chi}\chi} \frac{Y_{f_i}}{Y_{f_i}^\text{e.q.}}\frac{Y_{\overline{f}_i}}{Y_{\overline{f}_i}^\text{e.q.}}-  \gamma_{\overline{\chi}\chi \rightarrow \overline{f}_i f_i}\frac{Y_\chi}{Y_\chi^\text{e.q.}}\frac{Y_{\overline{\chi}}}{Y_{\overline{\chi}}^\text{e.q.}}\\
    &\simeq  \sum_i  \gamma_{\overline{f}_i f_i\rightarrow \overline{\chi}\chi},\label{eq:Boltz3}
\end{align}
where we applied the freeze-in approximation in the last step and for simplicity we  compute the scattering rate densities via Maxwell Boltzmann-averaging \cite{Gondolo:1990dk,Davidson:2008bu}  
\begin{align}\label{eq:av1}
    \gamma\left(a+b\rightarrow i+j+\dots \right) &=  \braket{\sigma \left|\vec{v}\right|} n_a^\text{eq.}n_b^\text{eq.}\\
    &=    \frac{  T}{32\pi^4}\int_{s_\text{min}}^\infty \text{d}s\; s^\frac{3}{2} \lambda\left(1,\frac{m_a^2}{s},\frac{m_b^2}{s}\right) K_1\left(\frac{\sqrt{s}}{T}\right)\; \sigma \nonumber
\end{align}
with
\begin{equation}
  \lambda(a,b,c) \equiv  (a-b-c)^2-4\; b c  \quad \text{and} \quad  s_\text{min} = \text{max}\left[\left(m_a+m_b\right)^2,\left(m_i+m_j+\dots\right)^2\right]
\end{equation}
instead of averaging with Fermi-Dirac statistics. By neglecting the masses of the DM and SM fermions the simpler Maxwell-Boltzmann average allows us to find an analytical expression by employing the relation \cite{Davidson:2008bu}
\begin{equation}
    \int_0^\infty \text{d}x\;  x^n K_1(x) = 2^{n-1} \Gamma\left(1+\frac{n}{2}\right) \Gamma\left(\frac{n}{2}\right)
\end{equation}
so that  
\begin{equation}\label{eq:gamma}
    \gamma\left(\overline{\chi}\chi \rightarrow \overline{f}_i f_i\right) =   \gamma\left( \overline{f}_i f_i \rightarrow \overline{\chi}\chi \right) =\frac{ 8}{ \pi^5} \alpha_{\chi i}  T^8.
\end{equation}
Note that while  the functional forms above are the same the densities depend on the different temperatures of the SM and DM baths. The fact that both densities are equal for equal temperatures reflects the principle of detailed balance, so that the right hand side of the Boltzmann equation vanishes in thermal equilibrium \cite{DAgnolo:2015ujb}. Owing to our previous simplifying assumptions we will only work with relativistic fermions in the SM plasma. Annihilations from non-relativistic fermions will be Boltzmann-suppressed at $T<m_{f_i}$ and therefore less important than relativistic processes.
From this we can deduce the more familiar interaction rate for relativistic SM fermions ($g_{f_i}=2$)
\begin{equation}\label{eq:int-rate}
    \Gamma\left( \overline{f}_i f_i \rightarrow \overline{\chi}\chi \right) = \frac{ \gamma\left( \overline{f}_i f_i \rightarrow \overline{\chi}\chi \right)}{n_{f_i}^\text{eq.}} = \frac{ 16}{ 3 \xi(3) \pi^3} \alpha_{\chi i}  T^5,
\end{equation}
which agrees with the estimate based on dimensional analysis for an effective four fermion operator that leads to  $\Gamma \sim  T^5 / v_\text{B-L}^4$. Our result is larger by only around 11\% compared to the result \cite{Heeck:2014zfa} found by averaging over Fermi-Dirac statistics and also using massless fermions. This numerical difference agrees with the findings of \cite{EscuderoAbenza:2020cmq} but we do not take percent level effects into account since what matters for freeze-in is the order of magnitude of the couplings and not their precise value.
In the effective operator limit the scattering rate is UV dominated so its maximum value is found at the largest available bath temperature after completion of reheating given by $T_\text{RH}$. As a consequence of our analysis for Higgs decays in \ref{sec:higgs} we will assume a reheating temperature $\SI{4}{\mega\electronvolt}\lesssim T_\text{RH}\lesssim \SI{5}{\giga\electronvolt}$. Since the SM fermions also couple to non-abelian gauge interactions the estimate \eqref{eq:therm-RH} is still approximately valid even though the SM fermions are not necessarily produced from inflaton decays. If we assume the inflaton decays to the SM like $h$, which definitely will be thermalized according to \eqref{eq:therm-RH}, and that $h$ decays or scatters to produce the SM fermions  it is plausible to expect a thermalized SM fermion bath immediately after reheating. Then in order to guarantee that we stay in the freeze-in regime the rate needs to satisfy
\begin{equation}
   \frac{ \sum_i  \Gamma\left( \overline{f}_i f_i \rightarrow \overline{\chi}\chi \right)}{H(T)} \Big|_{T=T_\text{RH}}<1
\end{equation} 
and we can use this to constrain the B-L breaking vev to be
\begin{equation}\label{eq:vev-cosmo}
    v_\text{B-L} \gtrsim \SI{56.8}{\tera\electronvolt} \cdot \left(\frac{T_\text{RH}}{\SI{1}{\giga\electronvolt}}\right)^\frac{3}{4} \cdot \left(\frac{\sum_i N_i (T_\text{RH}) }{11.67}\right)^\frac{1}{4} \cdot \left(\frac{76}{g_{*\rho}(T_\text{RH})}\right)^\frac{1}{8},
\end{equation}
which is a stronger constraint than the laboratory bound \eqref{eq:vevbound}.
For the above we summed over all the relativistic fermions because of the sum on the right hand side of 
\eqref{eq:Boltz2}. Moreover we used that for all SM leptons $Q_l^2=1$, for quarks $Q_q^2 = \frac{1}{9}$ with $N_c=3$ colors and assumed all leptons and quarks except the top and bottom quark to be relativistic at $T_\text{RH}=\SI{1}{\giga\electronvolt}$. We compute the effective coupling of the relativistic SM fermions as 
\begin{align}\label{eq:effCoupl}
    \sum_i N_i &\equiv \sum_i \left(N_c\right)_i \left(Q\left(f_{i\;L}\right)^2+Q\left(f_{i\;R}\right)^2\right)\\
    &= 3+ 2\sum_{l=e,\mu,\tau} \theta\left(T-\frac{m_l}{3}\right)+ 2 \theta\left(T-\SI{200}{\mega\electronvolt}\right) + \frac{2}{3} \sum_{q=t,b,c}\theta\left(T-\frac{m_q}{3}\right) \nonumber.
\end{align}
Here we treat a fermions as relativistic as long as $E\simeq 3 T > m_f$. In the above definition the first 3 stands for the SM neutrinos and the contribution from the charged leptons and quarks is multiplied by a 2 because both chiralities produce DM. We only wrote out the contributions from the heavy quarks explicitly and the term $ 2 \theta\left(T-\SI{200}{\mega\electronvolt}\right)$ is the contribution from $u,d,s$, whose mass is below the temperature of the QCD phase transition at $T_\text{QCD}\simeq \SI{200}{\mega\electronvolt}$. Below this transition all quarks hadronize and at least for a short period of temperature the light  mesons are still relativistic and should be taken into account \cite{Barger:2003zh}. The inclusion of these particles requires the use of form-factors and we ignore them because they quickly become non-relativistic and hence the rate density becomes double-Boltzmann suppressed compared to the contributions from $\nu_l$ and $e^-$. Note that we can reuse this estimate to make sure that the same interaction does not equilibrate the $\nu_R$ with charge $Q_1 = -2$: The cross section \eqref{eq:xsec} also applies to $\nu_R$ by replacing 
\begin{equation}\label{eq:repl}
    Q\left(\chi_L\right)^2+Q\left(\chi_R\right)^2 = 9 \quad \text{with}\quad  Q_1^2 = 4
\end{equation}
which is valid for both possible DM charge assignments \eqref{eq:irr1} and \eqref{eq:irr2}. Therefore the $\nu_R$ production rate is always smaller than the DM production rate so that \eqref{eq:vev-cosmo} ensures that there is no thermal population of $\nu_R$. We proceed by analytically solving \eqref{eq:Boltz3}
\begin{equation}\label{eq:DMyield}
    Y_\chi (z_0) =\mathcal{C}_\text{DM} \int_{z_\text{RH}=1}^{z_0=\infty} \text{d}z\; \frac{ \sum_i N_i(z)}{g_{*S}(z) \sqrt{g_{*\rho}(z)}} \frac{1}{z^4} \quad \text{with} \quad \mathcal{C}_\text{DM} = 0.32  \frac{M_\text{Pl} T_\text{RH}^3}{v_\text{B-L}^4}.
\end{equation}
Here the reheating temperature $T_\text{RH}$ acts as a UV-regulator for the effective cross section and if we were to consider $T_\text{RH}\rightarrow \infty$ we would need to use the full kinematic dependence of the $Z'$ propagator to unitarize the rate.
For the estimate we again replace the relativistic number of d.o.f with their values at $T_\text{RH}$ (see \ref{sec:higgs}) so that \eqref{eq:conv} evaluates to
\begin{align}\label{eq:relic-abun}
      \Omega_\text{DM} h^2 \simeq 0.12 &\cdot \left(\frac{m_\text{DM}}{\SI{10}{\kilo\electronvolt}}\right) \cdot \left(\frac{T_\text{RH}}{\SI{1}{\giga\electronvolt}}\right)^3 \cdot \left(\frac{\SI{172}{\tera\electronvolt}}{v_\text{B-L}}\right)^4\\
     & \cdot \left(\frac{\sum_i N_i(T_\text{RH})}{11.67}\right) \cdot  \left(\frac{76}{g_{*S}(T_\text{RH})}\right) \cdot \sqrt{\frac{76}{g_{* \rho}(T_\text{RH})}} \nonumber.
\end{align}
Note that one can not set $m_\text{DM}$ to arbitrarily large values since we neglected the phase space suppression for the finite DM mass in \eqref{eq:xsec}. As a rule of thumb our results apply as long as $m_\text{DM}\lesssim T_\text{RH}$. For the numerical evaluation we use the full temperature dependence of $g_{*S}$ and $g_{*\rho}$ by employing the fitting functions from \cite{Wantz:2009it}, which agree up to less than one percent with the exact expressions except during the QCD phase transition and during $e^+ e^-$ annihilations, where the differences are about four percent. Figures \ref{fig:YDM1} and \ref{fig:YDM2} illustrate the behaviour of the DM abundance today for different values of the reheating temperature, DM mass and $v_\text{B-L}$ together with the observed relic abundance \protect\cite{Planck:2018vyg}. As previously alluded to one can see that the yield reaches its asymptotic value shortly after reheating and its final value strongly depends on $T_\text{RH}$ as expected for UV freeze-in. 
\newline
Before closing we would like to emphasize that the SM like Higgs can also mediate SM fermions annihilating to DM via the effective interaction in \eqref{eq:EFT}. The corresponding scattering rate density is found from \eqref{eq:gamma}
\begin{equation}
  \sum_{f_i}  \gamma^h \left( \overline{f}_i f_i \rightarrow \overline{\chi}\chi \right)  \simeq \frac{12}{\pi^5} \left(\frac{m_\text{DM}}{v_H}\right)^2 \sum_{f_i}  \left(\frac{m_{f_i}^\text{eff.}(T)}{v_H}\right)^2 T^8,
\end{equation}
where 
\begin{align}\label{eq:masssum}
    \sum_{f_i} \left(m_{f_i}^\text{eff.}(T)\right)^2 &\equiv  \sum_{l=e,\mu,\tau} m_l^2\; \theta\left(T-\frac{m_l}{3}\right)
    + 3  \left(m_u^2+m_d^2+m_s^2\right)\; \theta\left(T-\SI{200}{\mega\electronvolt}\right)\\
   &+  3 \sum_{q=t,b,c}  m_q^2\; \theta\left(T-\frac{m_q}{3}\right)\nonumber
\end{align}
encodes the couplings of the relativistic SM fermions to the Higgs in analogy with \eqref{eq:effCoupl}. We neglect the coupling to the active neutrinos as it scales with their mass. The Higgs mediated interaction does not thermalize at reheating provided that
\begin{equation}
    m_\text{DM} \lesssim \SI{1}{\giga\electronvolt}\cdot \left(\frac{g_{*\rho}(T_\text{RH})}{76}\right)^\frac{1}{4} \cdot \left(\frac{\SI{1}{\giga\electronvolt}}{T_\text{RH}}\right)^\frac{3}{2},
\end{equation}
which is stronger than the bound from invisible Higgs decays \eqref{eq:DMBound1}.
The estimate for the Higgs mediated relic abundance is straightforward and by comparing with \eqref{eq:DMyield} we arrive at
\begin{equation}\label{eq:ratio}
    \frac{\Omega_\text{DM}^h}{\Omega_\text{DM}^{Z'}} \simeq \frac{3}{2} \left(\frac{v_\text{B-L}}{m_h}\right)^4 \left(\frac{m_\text{DM}}{v_H}\right)^2 \frac{ \sum_{f_i} \left(\frac{m_{f_i}^\text{eff.}(T_\text{RH})}{v_H}\right)^2}{\sum_i N_i(T_\text{RH})}.
\end{equation}
If we demand that this additional contribution is smaller than the $Z'$ mediated one we obtain an upper limit on the DM mass of
\begin{equation}\label{eq.bound}
    m_\text{DM}\lesssim \SI{3}{\mega\electronvolt}\cdot \sqrt{\frac{\Omega_\text{DM}^h / \Omega_\text{DM}^{Z'}}{1\%}} \cdot \sqrt{\frac{\sum_i N_i(T_\text{RH})}{11.67}} \cdot \left(\frac{\SI{172}{\tera\electronvolt}}{v_\text{B-L}}\right)^2,
\end{equation}
which was evaluated at $T_\text{RH} = \SI{1}{\giga\electronvolt}$, where all charged fermions except the top and bottom quark contribute. This represents the strongest upper bound on the DM mass and is the reason why we only consider DM with typical masses at the keV-scale. We depict contours in the $T_\text{RH}$ versus $v_\text{B-L}$ plane that reproduce the measured DM abundance today for multiple  representative masses that agree with the Lyman-$\alpha$ bound and \eqref{eq.bound} in figure \ref{fig:paramspaceGauge}.

\subsection{Dark matter phenomenology}
\noindent Owing to our choice of $\mathcal{Z}_5$ symmetry the DM is absolutely stable and does not mix with the SM neutrinos. Consequently the DM has no radiative decay mode to a $\nu_L$ plus a photon. This decay constitutes the canonical signature of keV scale sterile neutrino DM that is being looked for via   X-ray searches investigating the diffuse X-Ray background or dwarf galaxies \cite{Dolgov:2000ew,Abazajian:2001vt,Boyarsky:2005us,2009arXiv0906.1788D} (see also \cite{Drewes:2016upu} and references therein).
\newline
The coupling to the $Z'$ and the Higgs induce velocity independent dark matter self interaction cross sections. However due to the small DM mass and $v_\text{B-L}\gg m_h\gg m_\text{DM}$ the resulting transfer cross sections \cite{Feng:2009hw,Tulin:2013teo,Kahlhoefer:2017umn} are far to small to help with the \enquote{cusp-core} and \enquote{too-big-to-fail}-problems \cite{Vogelsberger:2012,Rocha:2012jg,Vogelsberger:2013,Peter:2012jh} or even to come into conflict with bounds from the Milky way or the Bullet cluster  \cite{Vogelsberger:2012,Rocha:2012jg,Peter:2012jh}. 
\newline
The aforementioned self interactions could lead to efficient DM self scatterings which would lead to kinetic equilibrium of the DM population in the early universe. Because of the separation of scales between $m_h$ and $v_\text{B-L}$ we only focus on the individual contributions and ignore the interference term. For the Higgs mediated interaction we find in the limit $s,\;m_\text{DM}\ll m_h$ 
\begin{equation}
    \sigma_h(\overline{\chi}\chi \rightarrow \overline{\chi}\chi) \simeq \frac{1}{\pi}\left(\frac{m_\text{DM}}{v_H}\right)^4 \frac{s}{m_h^4}
\end{equation}
and use the methods of section \ref{sec:gauge1} to compute
\begin{equation}
    \gamma_h(\overline{\chi}\chi \rightarrow \overline{\chi}\chi)\simeq \frac{24}{\pi^5} \left(\frac{m_\text{DM}}{v_H}\right)^4  \frac{T^8}{m_h^4}.
\end{equation}
By comparing the interaction rate $\Gamma_h= \gamma_h/ n_\chi$, where $n_\chi$ is the DM number density with $g_\chi=4$ degrees of freedom,  to the Hubble rate evaluated at reheating we find that the DM is not in kinetic equilibrium with itself at $T_\text{RH}$ as long as
\begin{equation}
    m_\text{DM} \lesssim \SI{475}{\mega\electronvolt} \cdot \left(\frac{\SI{5}{\giga\electronvolt}}{T_\text{RH}}\right)^\frac{3}{4} \cdot \left(\frac{g_{*\rho}(T_\text{RH})}{85}\right)^\frac{1}{8}.
\end{equation}
Similarly to find $\Gamma_{Z'}(\overline{\chi}\chi \rightarrow \overline{\chi}\chi)$ we can reuse the result \eqref{eq:int-rate} by replacing the charges (see \eqref{eq:effCoupl})
\begin{equation}
    \sum_i N_i \quad \text{with} \quad Q\left(\chi_L\right)^2+Q\left(\chi_R\right)^2 = 9.
\end{equation}
The $Z'$ mediated diagram does not equilibrate  the DM with itself at reheating provided that
\begin{equation}
    v_\text{B-L} \gtrsim \SI{175}{\tera\electronvolt}\cdot \left(\frac{T_\text{RH}}{\SI{5}{\giga\electronvolt}}\right)^\frac{3}{4} \cdot \left(\frac{85}{g_{*\rho}(T_\text{RH})}\right)^\frac{1}{8}.
\end{equation}
We conclude that scattering can lead to kinetic equilibrium of the DM at early times for certain choices of parameters. Since both rates arise from effective operators they decrease with temperature, which means that even if the DM was thermalized with itself initially it will fall out of kinetic equilibrium during the evolution of the universe.
\newline
As a consequence of the constraint \eqref{eq:DMBound1} we only investigate very light DM with typical masses below $\SI{2}{\giga\electronvolt}$. Since nuclear recoil experiments basically have no sensitivity for sub-GeV DM due to kinematics, there has been a growing interest in studying atomic bound state  electrons as targets for direct detection of light DM \cite{Essig:2017kqs}. In order to estimate whether these targets can be used to find our DM candidate, we compute the Higgs and $Z'$ mediated cross sections for non-relativistic DM in the electron rest frame and expand to leading order in $v_\text{DM}\ll1$:
\begin{align}
    \sigma_h(\chi e^-\rightarrow \chi e^-) &\simeq \frac{16}{\pi} \left(\frac{m_e m_\text{DM}}{m_e + m_\text{DM}}\right)^6 \frac{v_\text{DM}^2}{m_h^4 v_H^4}\\
    &\simeq \begin{cases}
            \SI{4e-71}{\centi\meter\squared}\cdot \left(\frac{v_\text{DM}}{10^{-3}}\right)^2\quad \text{for}\quad m_\text{DM} \gg m_e\\
            \SI{2e-81}{\centi\meter\squared} \cdot \left(\frac{m_\text{DM}}{\SI{10}{\kilo\electronvolt}}\right)^6 \cdot  \left(\frac{v_\text{DM}}{10^{-3}}\right)^2 \quad \text{for}\quad m_\text{DM} \ll m_e \nonumber
        \end{cases}\\
    \sigma_{Z'}(\chi e^-\rightarrow \chi e^-) &\simeq \frac{4}{\pi} \frac{m_e^4 m_\text{DM}^4}{(m_e + m_\text{DM})^6}  \frac{\left(Q\left(\chi_L\right)+Q\left(\chi_R\right)\right)^2}{v_\text{B-L}^4}v_\text{DM}^2\\
    &\simeq \begin{cases}
            \SI{6.5e-66}{\centi\meter\squared}  \cdot \left(\frac{\SI{10}{\mega\electronvolt}}{m_\text{DM} }\right)^2 \cdot  \left(\frac{v_\text{DM}}{10^{-3}}\right)^2  \cdot  \left(\frac{\SI{967}{\tera\electronvolt}}{v_\text{B-L}}\right)^4\quad \text{for}\quad m_\text{DM} \gg m_e\\
            \SI{3.7e-67}{\centi\meter\squared}\cdot \left(\frac{m_\text{DM}}{\SI{10}{\kilo\electronvolt}}\right)^4\cdot \left(\frac{v_\text{DM}}{10^{-3}}\right)^2 \cdot  \left(\frac{\SI{172}{\tera\electronvolt}}{v_\text{B-L}}\right)^4\quad \text{for}\quad m_\text{DM} \ll m_e \nonumber
        \end{cases}
\end{align}
The Higgs mediated cross section comes with two more powers of both $m_e$ and $m_\text{DM}$ compared to the $Z'$ mediated one, because the couplings to the Higgs are proportional to the aforementioned masses.
In the above we chose $v_\text{B-L}$ to reproduce the observed DM relic density for a given DM mass.
The best current limit including form factors for bound state electrons is $\sigma\lesssim 10^{-40}\SI{}{\centi\meter\squared}$ \cite{XENON:2019gfn,XENON:2021myl}. One can see that direct detection via electrons is not a viable search strategy for our DM candidate owing to the small values of $m_\text{DM}$ and the large $v_\text{B-L}$ necessary for freeze-in.
 
\section{Dark Radiation}\label{sec:darkRad}

\noindent 
The SM prediction for the number of relativistic neutrinos is \cite{Gnedin:1997vn,Mangano:2005cc,deSalas:2016ztq,Froustey:2020mcq,Akita:2020szl,Bennett:2020zkv}
\begin{equation}
    N_\text{eff.} = 3.0440 \pm 0.0002,
\end{equation}
and the small deviation from the value of $3$ expected for three generations of $\nu_L$ comes from the fact that their decoupling from the SM bath is not instantaneous. Additional relativistic degrees of freedom are usually referred to as dark radiation (DR).
From the observed abundance of light elements produced during Big Bang Nucleosynthesis (BBN) one infers $N_\text{eff.}^\text{BBN} = 2.95^{+0.56}_{-0.52}$  \cite{Planck:2018vyg}. Combined analyses of the current Planck CMB data together with  Baryon Acoustic oscillations (BAO) found $ N_\text{eff.}^\text{Planck+BAO} = 2.99^{+0.34}_{-0.33}$ \cite{Planck:2018vyg}. This can be recast as 
\begin{equation}
    \Delta N_\text{eff.}^\text{Planck+BAO} \simeq 0.28\;@\; 2\sigma \; \text{C.L.}\;.
\end{equation}
Currently there is a lot experimental effort to improve this bound:
The South Pole Telescope \cite{SPT-3G:2014dbx} and the Simons observatory \cite{SimonsObservatory:2018koc} both aim to reach  $\Delta N_\text{eff.} \lesssim 0.12 \;@\; 2\sigma \; \text{C.L.}$ while the upcoming CMB Stage 4 (CMB-S4), experiment \cite{Abazajian:2019eic,annurev-nucl-102014-021908,CMB-S4:2016ple} and NASA's PICO proposal \cite{NASAPICO:2019thw} have a sensitivity forecast of  $\Delta N_\text{eff.} =0.06\;@\; 2\sigma \; \text{C.L}$. There is also the planned  CORE experiment by the ESA \cite{CORE:2017oje} with similar goals.

\subsection{Dark Matter as dark radiation}
\noindent Since the dark matter is out of equilibrium with the SM bath, its typical momentum after production   can in principle be vastly different from the temperature of the SM. Even though the DM is non-relativistic today, it might have been relativistic at the time of BBN or CMB decoupling. One can find a condition for having a relativistic DM particle at the SM bath temperature $T$ \cite{Decant:2021mhj}
\begin{equation}
    m_\text{DM} < \frac{T_\gamma(t_0) \left(\frac{g_{*S}\left(T_\gamma(t_0)\right)}{g_{*S}(T_\text{RH})}\right)^\frac{1}{3}}{a(T)} = \frac{\SI{2e-7}{\kilo\electronvolt} \cdot \left(\frac{100}{g_{*S}(T_\text{RH})}\right)^\frac{1}{3}}{a(T)}.
\end{equation}
Here $T_\gamma(t_0)$ is the photon temperature today and $\frac{g_{*S}\left(T_\gamma(t_0)\right)}{g_{*S}(T_\text{RH})}$ is the ratio in the number of relativistic degrees of freedom in entropy today versus the number at the time of DM production, which we approximate with $T_\text{RH}$. $a(T)$ denotes the scale factor, whose value ranges from $\simeq 10^{-10}$ at the time of BBN ($T\simeq \SI{1}{\mega\electronvolt}$) to $\simeq 10^3$ at the time of CMB decoupling ($T\simeq \SI{1}{\electronvolt}$). Consequently our DM candidate can only be relativistic around BBN, but not at recombination. The contribution of the DM to $\Delta N_\text{eff.}$ at BBN temperatures was found to be \cite{Decant:2021mhj,Li:2021okx}
\begin{equation}
    \Delta N_\text{eff.}(T_\text{BBN}) \simeq 3.4\times 10^{-4} \cdot \left(\frac{\Omega_\text{DM}h^2}{0.12}\right) \cdot \left(\frac{\SI{10}{\kilo\electronvolt}}{m_\text{DM}}\right)\cdot \left(\frac{100}{g_{*S}(T_\text{RH})}\right)^\frac{1}{3}
\end{equation}
and is negligible compared to the expected sensitivities. Note that the above estimate relied on the FIMP being produced from a decay, however we do not expect production from scattering to significantly alter the order of magnitude of the result.

\subsection{Right handed neutrinos as dark radiation}\label{sec:gauge2}
\noindent 
Due to their feeble effective Yukawa interaction with the left handed neutrinos (see \eqref{eq:numass}) the $\nu_R$ never equilibrate with the SM \cite{Dick:1999je} and the freeze-in   of the aforementioned interaction contributes an even more negligible amount of \cite{Luo:2020fdt}
\begin{equation}
    \Delta N_\text{eff.} \simeq 7.5\times10^{-12}\;\cdot \left(\frac{m_\nu}{\SI{0.1}{\electronvolt}}\right)^2
\end{equation}
in standard Big Bang cosmology.
Gauge annihilations of SM fermions via the $Z'$ can also create $\nu_R$. From section \ref{sec:gauge1} we already know that  if we want to produce the DM from freeze-in the $\nu_R$ production will occur in the freeze-in regime as well. The corresponding cross section is given by \eqref{eq:xsec} under the replacement \eqref{eq:repl} and $\alpha_{\chi i}\rightarrow \alpha_{\nu_R i}$. We can write down the coupled Boltzmann equations for the evolution of the SM and DM energy densities \cite{Luo:2020sho}
\begin{align}
    \frac{\text{d}\rho_\text{SM}}{\text{d}t}+3 H \left(\rho_\text{SM}+P_\text{SM}\right) &= -C_{\rho},\label{eq:number1}\\
    \frac{\text{d}\rho_{\nu_R}}{\text{d}t}+3 H \left(\rho_{\nu_R}+P_{\nu_R}\right) &= C_{\rho} \label{eq:number2},
\end{align}
where $P$ denotes the pressure density. 
Adding both Boltzmann equations gives the result expected from the continuum equation
\begin{equation}
  \sum_{i=\text{SM},\nu_R}  \frac{\text{d}\rho_i}{\text{d}t}+3 H \left(\rho_i+P_i\right) = 0.
\end{equation}
Making use of the equation of state for radiation allows us to write
\begin{equation}
    \rho_i + P_i = \frac{4}{3}\rho_i,\quad i=\text{SM},\nu_R.
\end{equation}
The right hand side of the Boltzmann equations is known as the collision term and parameterizes the energy exchange between the SM and DM baths. It can be written as 
\cite{EscuderoAbenza:2020cmq,Biswas:2021kio}
\begin{align}
     C_{\rho} &=   \sum_i \braket{E \sigma \left|\vec{v}\right|}_{\overline{f_i}f_i \rightarrow \overline{\nu_R}\nu_R} n_{f_i} n_{\overline{f}_i}-\braket{E \sigma \left|\vec{v}\right|}_{   \overline{\nu_R}\nu_R\rightarrow \overline{f_i}f_i} n_{\nu_R} n_{\overline{\nu}_R},\\
     & \simeq   \sum_i \braket{E \sigma \left|\vec{v}\right|}_{\overline{f_i}f_i \rightarrow \overline{\nu_R}\nu_R} n_{f_i}^\text{eq.} n_{\overline{f}_i}^\text{eq.},
\end{align}
where we neglect the back-reaction from the $\nu_R$ bath in the freeze-in approximation in the second line. The quantities $\braket{E \sigma \left|\vec{v}\right|}$ are functions of the respective bath temperatures and are defined completely analogous  to $\braket{ \sigma \left|\vec{v}\right|}$ in \eqref{eq:av1} as \cite{Gondolo:1990dk,EscuderoAbenza:2020cmq,Biswas:2021kio} 
\begin{align}
    \delta\left(a+b\rightarrow i+j+\dots \right) &=  \braket{E \sigma \left|\vec{v}\right|} n_a^\text{eq.}n_b^\text{eq.}\\
    &=
    \frac{  T}{64\pi^4}\int_{s_\text{min}}^\infty \text{d}s\; s^2 \lambda\left(1,\frac{m_a^2}{s},\frac{m_b^2}{s}\right)\left(1+\frac{m_a^2-m_b^2}{2}\right) K_2\left(\frac{\sqrt{s}}{T}\right)\; \sigma \nonumber.
\end{align}
$K_2$ is the modified Bessel function of the second kind, which arises compared to the $K_1$ in $\braket{ \sigma \left|\vec{v}\right|}$ due to the presence of a factor of $E$ in the thermal average. Again we use Maxwell-Boltzmann statistics instead of the correct Fermi-Dirac averaging to obtain simpler analytic results. By employing the relation
\begin{equation}
    \int_0^\infty \text{d}x\; \text{K}_2(x) = 2^{n-1} \Gamma\left(\frac{n-1}{2}\right)  \Gamma\left(\frac{n+3}{2}\right)\quad \text{for}\quad n>1 
\end{equation}
we can compute the average for massless initial and final states
\begin{equation}
    \delta(\overline{f_i}f_i \rightarrow \overline{\nu_R}\nu_R)=  \delta( \overline{\nu_R}\nu_R \rightarrow \overline{f_i}f_i )  = \frac{8}{\pi^5} \alpha_{\nu_R} T^9.
\end{equation}
Note again that in the above one has to take into account that the rate densities depend on the different bath temperatures. The scaling of this energy exchange rate density is consistent with dimensional analysis as it scales like the  rate density \eqref{eq:gamma} for the DM abundance multiplied by another factor of $T$. Since we do not know the phase-space distribution function of the non-thermal $\nu_R$ we do not know their temperature so we compute their energy density directly from solving the Boltzmann equation.
If we neglect the energy loss of the SM bath, which is the basis of the freeze-in scenario and assume that the SM entropy is conserved we  find 
\cite{Luo:2020fdt}
\begin{equation}\label{eq:rho}
    \rho_{\nu_R}(T) \simeq 2\cdot s_\text{SM}(T)^\frac{4}{3} \int^{T_\text{RH}}_T \text{d}\tilde{T}\; \frac{s_\text{SM}'(\tilde{T})}{3 s_\text{SM}(\tilde{T})^\frac{7}{3} H(\tilde{T})} \delta_{\overline{f_i}f_i \rightarrow \overline{\nu_R}\nu_R}(\tilde{T})
\end{equation}
in terms of the SM temperature $T$ and use this to compute \cite{Luo:2020fdt}
\begin{equation}\label{eq:neff}
    \Delta N_\text{eff.}(T) =  2\cdot\frac{4}{7}\; g_{*\rho}(T) \left(\frac{10.75}{g_{*S}(T)}\right)^\frac{4}{3}  \frac{\rho_{\nu_R}(T)}{\rho_\text{SM}(T)}\quad \text{with} \quad \rho_\text{SM}(T) = \frac{\pi^2}{30} g_{*\rho}(T) T^4,
\end{equation}
where the first  factor of two in \eqref{eq:rho} accounts for the fact that the  $\nu_R$ have $g_{\nu_R}=2$ spin polarizations and the second one in \eqref{eq:neff} for two generations of $\nu_R$. 
At temperatures below the electron mass $e^+ e^-$ annihilations heat the SM plasma compared to the decoupled species so that by using $ g_{*S}(T<m_e) = \frac{43}{11}$ we recover the more familiar formula
\begin{equation}
    \Delta N_\text{eff.} (T<m_e) = 2\cdot\frac{8}{7} \left(\frac{11}{4}\right)^\frac{4}{3} \frac{\rho_{\nu_R}(T)}{\rho_\gamma (T)}.
\end{equation}
For the regime where the $\nu_R$ were initially in thermal equilibrium with the SM  until they decoupled at $T_\text{FO}$  before the $\nu_L$ decoupling one  would find \cite{Planck:2018vyg,CMB-S4:2016ple}
\begin{equation}\label{eq:neff-FO}
    \Delta N_\text{eff.}^\text{eq.} =2\cdot \frac{g_{\nu_R}}{2} \left(\frac{10.75}{g_{*S}(T_\text{FO})}\right)^\frac{4}{3}
\end{equation}
instead.
Integrating the collision term in \eqref{eq:rho}  is straightforward and we find
\begin{equation}\label{eq:DRyield}
    \rho_{\nu_R} (T) =\mathcal{C}_{\nu_R}(T) \int^{T_\text{RH}}_{T} \text{d}\tilde{T}\; \frac{ \sum_i N_i(\tilde{T})\; \tilde{T}^2}{g_{*S}(\tilde{T})^\frac{4}{3} \sqrt{g_{*\rho}(\tilde{T})}} 
\end{equation}
with
\begin{equation}
    \mathcal{C}_{\nu_R}(T) = 0.13\;g_{*S}(T)^\frac{4}{3} \frac{M_\text{Pl} T^4}{v_\text{B-L}^4}.
\end{equation}
Our estimate for the additional number of relativistic species is in the limit $T_\text{RH}\gg T$
\begin{align}
    \Delta N_\text{eff.}\simeq  1.6\times 10^{-4} &\cdot    \left(\frac{T_\text{RH}}{\SI{1}{\giga\electronvolt}}\right)^3 \cdot \left(\frac{\SI{172}{\tera\electronvolt}}{v_\text{B-L}}\right)^4 \\
    &\cdot \left(\frac{\sum_i N_i(T_\text{RH})}{11.67}\right) \cdot  \left(\frac{76}{g_{*S}(T_\text{RH})}\right)^\frac{4}{3} \cdot \sqrt{\frac{76}{g_{* \rho}(T_\text{RH})}}\nonumber.
\end{align}
As expected the abundance of non-thermal DR strongly depends on their production temperature $T_\text{RH}$. Note that while it seems that the above expression can lead to arbitrarily large values of $ \Delta N_\text{eff.}$ one should keep in mind, that the present treatment relying on \eqref{eq:rho} breaks down as soon as one starts to violate \eqref{eq:vev-cosmo} because the $\nu_R$ thermalize. In that case one can use \eqref{eq:neff-FO} to compute $ \Delta N_\text{eff.}$ from the freeze-out temperature and finds that it asymptotes to a value of two for two $\nu_R$. By plugging in the lower limit on $v_\text{B-L}$ from the DM production being out of thermal equilibrium in \eqref{eq:vev-cosmo} we find that the freeze-in contribution of $\nu_R$ via $Z'$ mediated scatterings is at least a factor of five below the sensitivities of the upcoming CMB experiments
\begin{equation}
    \Delta N_\text{eff.} < 1.2\times 10^{-2} \cdot \left(\frac{85}{g_{*S}(T_\text{RH}=\SI{5}{\giga\electronvolt})}\right)^\frac{4}{3}.
\end{equation}

\begin{figure}[!t]
\centering
    \begin{minipage}{0.75\textwidth}
    \includegraphics[width=\textwidth]{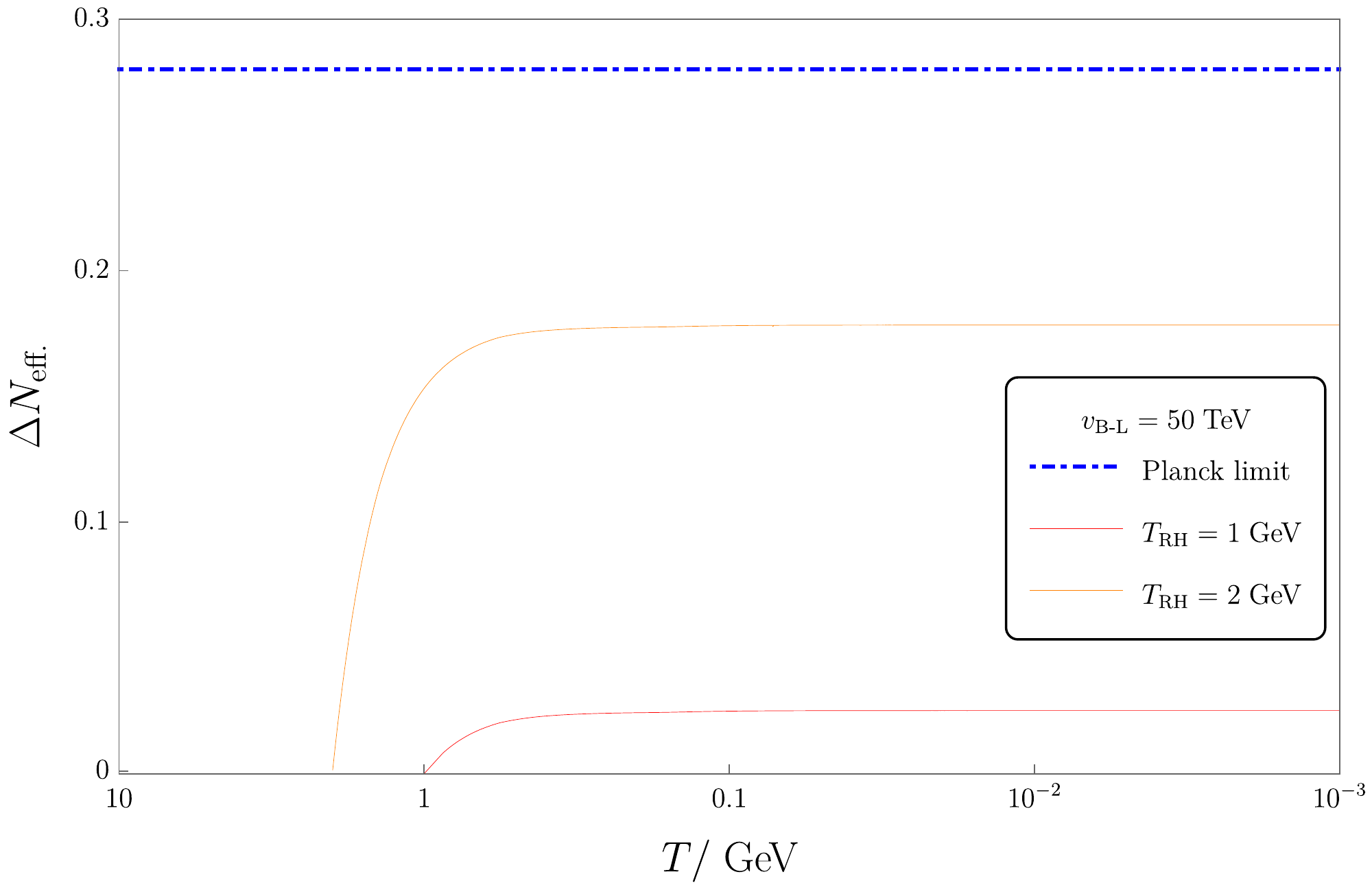}
    \caption{$\Delta N_\text{eff.}$  as a function of temperature for fixed $v_\text{B-L}$ and two different $T_\text{RH}$.}
    \label{fig:Neff1}
    \end{minipage}
        
    \vspace{0.05\textwidth}
    \begin{minipage}{0.75\textwidth}
    \includegraphics[width=\textwidth]{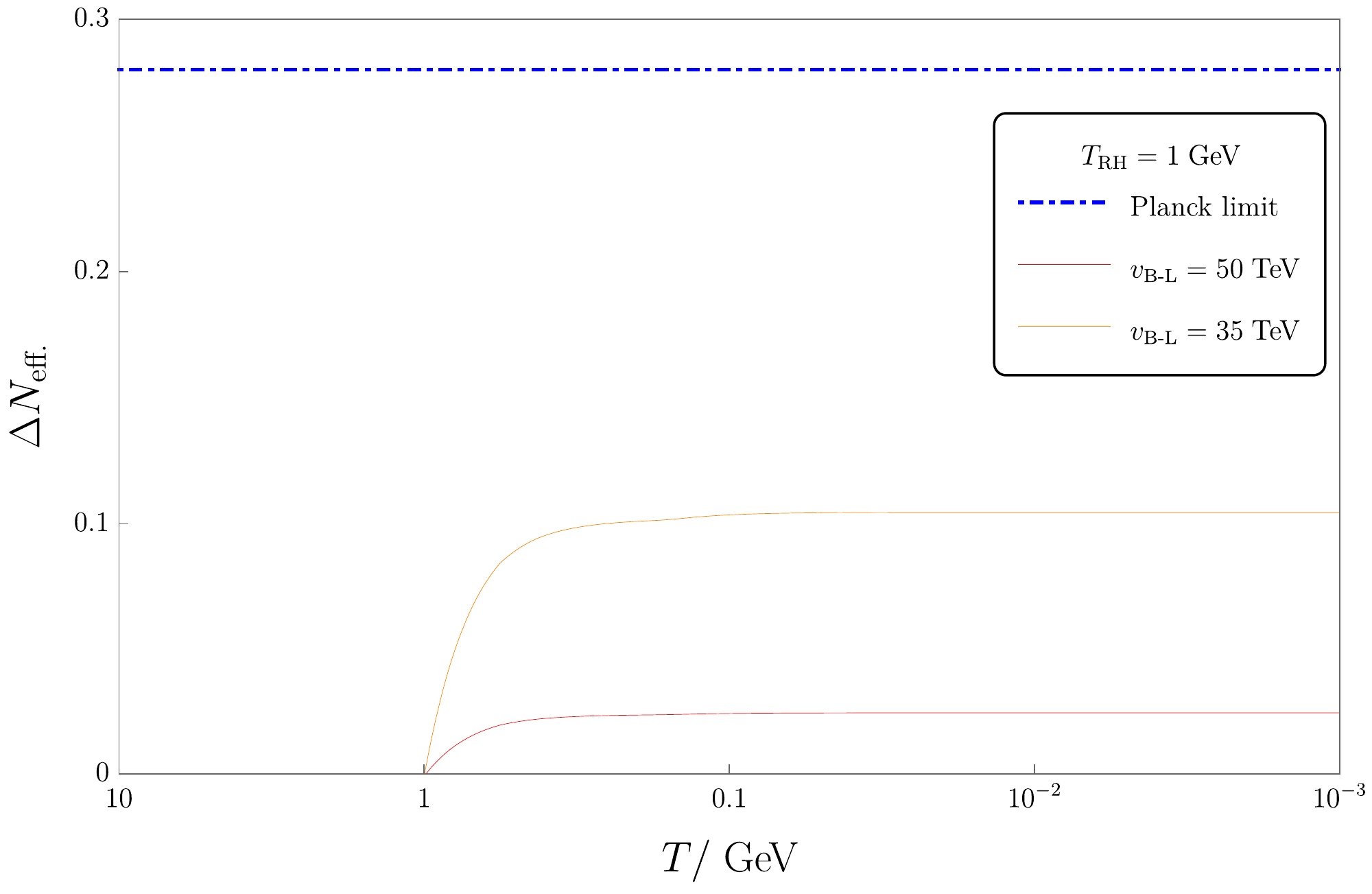}
    \caption{$\Delta N_\text{eff.}$ as a function of temperature for fixed $T_\text{RH}$ and two different   $ v_\text{B-L}$.}
    \label{fig:Neff2}
    \end{minipage}
\end{figure}

\begin{figure}[!t]
    \centering
    \includegraphics[width=0.8\textwidth]{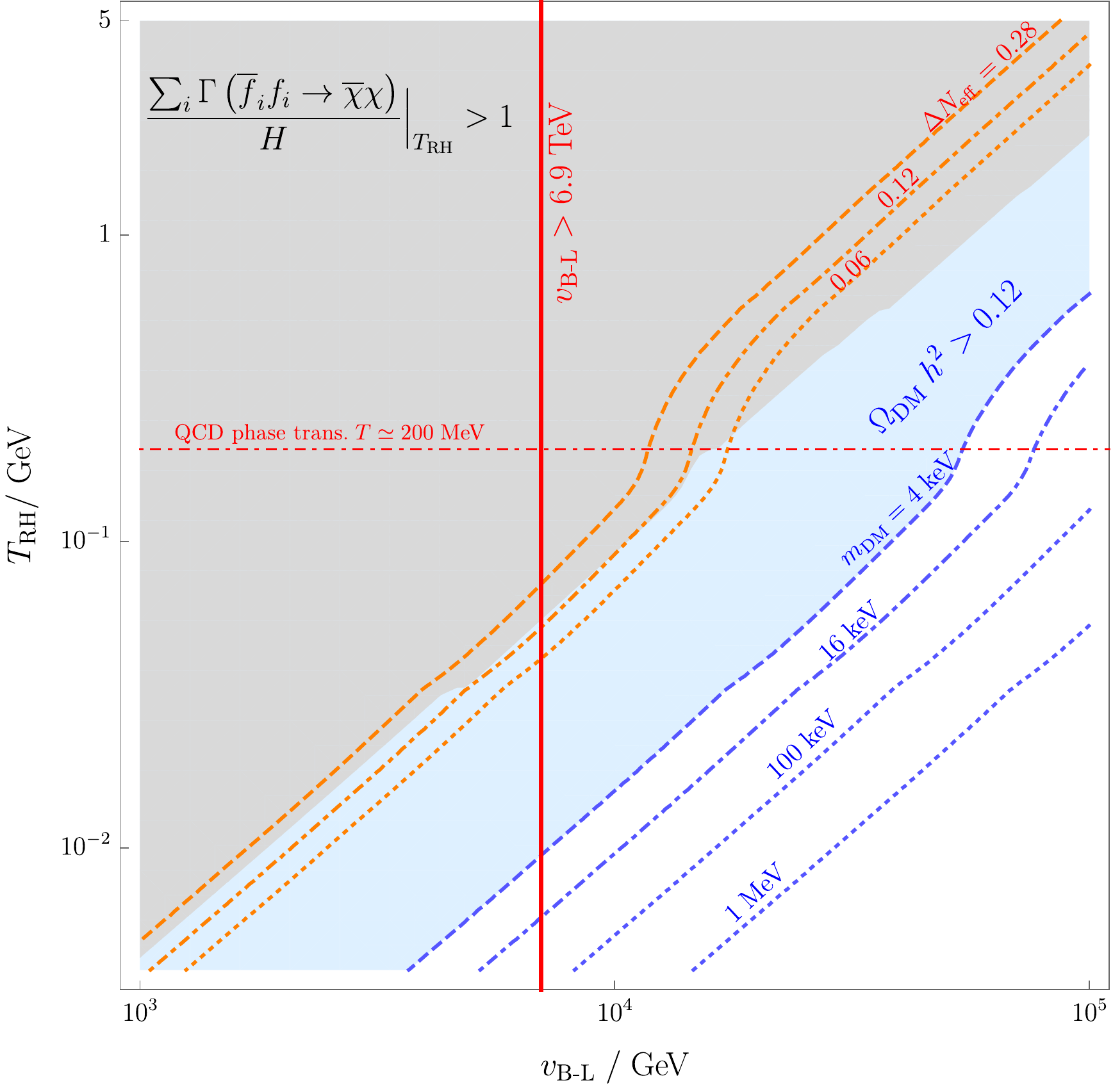}
    \caption{We depict the allowed combinations of the reheating temperature $T_\text{RH}$ and the scale of B-L breaking $v_\text{B-L}$. The blue shaded area indicates where DM would overclose the universe and the blue contours reproduce the observed DM relic density for $m_\text{DM}\in\left[4,16,100,10^3\right]\text{keV}$. Furthermore we show the contours for generating  $\Delta N_\text{eff.}$ within the Planck bound  \protect\cite{Planck:2018vyg},  the estimated sensitivities of the  South Pole Telescope \protect\cite{SPT-3G:2014dbx}, the Simons observatory \protect\cite{SimonsObservatory:2018koc} and for the  CMB stage 4 experiment \protect\cite{Abazajian:2019eic,annurev-nucl-102014-021908,CMB-S4:2016ple} as well as PICO \protect\cite{NASAPICO:2019thw} . The grey area is excluded because the interaction producing DM would equilibrate see \protect\eqref{eq:vev-cosmo} and searches fom LEP  exclude $v_\text{B-L}<\SI{6.9}{\tera\electronvolt}$ \protect\cite{PhysRevD.70.093009}. }
    \label{fig:paramspaceGauge}
\end{figure}

\noindent We conclude that the interplay of the tiny rates $\sim v_\text{B-L}^{-4}$  together with the fact that we consider a  cosmology with a low reheating temperature  reduces the impact of $\nu_R$ and $\chi$ on $  \Delta N_\text{eff.}$ below all current and future sensitivities.
This opens up an interesting indirect way to test our model: Should observations ever point to  
$\Delta N_\text{eff.}>0.012$ our scenario for DM production is excluded.
\newline
For the numerical evaluation of \eqref{eq:rho} we proceed as in section \ref{sec:gauge1}.
The temperature dependence of $\Delta N_\text{eff.}$ was depicted in \ref{fig:Neff1} and \ref{fig:Neff2} together with the limit from Planck \protect\cite{Planck:2018vyg}.
For better visibility of the final DR yield we chose values of $v_\text{B-L}$ below the bound  \eqref{eq:vev-cosmo}. The curves in \ref{fig:Neff1}  demonstrate that the abundance strongly depends on the reheating temperature and  \ref{fig:Neff2} that it decreases with growing $v_\text{B-L}$. Both plots show how the final yield is reached shortly after reheating as was the case for DM production.
\newline
There is also a contribution to the annihilations of SM fermions to $\nu_R$ via the exchange of an SM like Higgs. The corresponding rate density reads in terms of the coupling  \eqref{eq:masssum}
\begin{equation}
       \sum_{f_i}    \delta^h (\overline{f_i}f_i \rightarrow \overline{\nu_R}\nu_R)\simeq \frac{12}{\pi^5} \left(\frac{m_\nu}{v_H}\right)^2   \sum_{f_i}  \left(\frac{m_{f_i}^\text{eff.}(T)}{v_H}\right)^2 T^9,
\end{equation}
and it does not thermalize at $T_\text{RH}$ due to the tiny coupling $\propto \left(m_\nu / v_H\right)^2$. The estimate for the ratio of the resulting DR yields is equal to \eqref{eq:ratio} under the replacement $m_\text{DM} \rightarrow m_\nu$. We find that we can neglect the freeze-in of $\Delta N_\text{eff.}$ via Higgs interactions as 
\begin{equation}
    \frac{\Delta N_\text{eff.}^h}{\Delta N_\text{eff.}^{Z'}}\simeq 10^{-17}\cdot \left(\frac{m_\nu}{\SI{0.1}{\electronvolt}}\right)^2 \cdot \left(\frac{v_\text{B-L}}{\SI{172}{\tera\electronvolt}}\right)^4 \cdot  \left(\frac{11.67}{\sum_i N_i(T_\text{RH})}\right).
\end{equation}
The above  was evaluated at $T_\text{RH} = \SI{1}{\giga\electronvolt}$, where all charged fermions except the top and bottom quark contribute.
Figure \ref{fig:paramspaceGauge} demonstrates the available parameter space for realizing the entire DM abundance from $\chi$s via freeze-in together with the predicted amount of dark radiation parameterized in terms of $\Delta N_\text{eff.}$. A few   comments are in order:
The gray region excluded by \eqref{eq:vev-cosmo} has a more rugged contour because of the sequence of Heaviside functions in the expression for the sum of fermion charges \eqref{eq:effCoupl}. Additionally there is a noticeable kink in all DM and DR contours, which occurs around the temperature of the QCD phase transition at  $T_\text{QCD} \simeq \SI{200}{\mega\electronvolt}$. The physical reason for this behaviour can be found by inspecting the expressions for the DM and DR yields in \eqref{eq:DMyield} and \eqref{eq:DRyield}:  The integrands in both cases depend on inverse powers of $g_{*S}(T)$ and $g_{*\rho}(T)$ and the number of relativistic degrees of freedom in entropy and energy both decrease drastically when the quarks and gluons confine at $T_\text{QCD}$. To keep the relic density or $\Delta N_\text{eff.}$ fixed one needs to compensate this increase of the integrand by allowing for a larger value of $v_\text{B-L}$, hence the contours appear to be shifted to the right below $T_\text{QCD}$, which is why for illustration we chose to display a straight line at the corresponding temperature. One should not forget that the factor of  $\sum_i N_i$ in both numerators also decreases sharply below $T_\text{QCD}$, but is approximately cancelled by one of the factors in the denominator leaving one factor in the denominator leading to the previously explained behaviour.
\newline
It is evident from  \ref{fig:paramspaceGauge} that the Planck constraint on $\Delta N_\text{eff.}$ would only be relevant for DM masses far below $\SI{4}{\kilo\electronvolt}$, which is already excluded by the Lyman-$\alpha$ constraints. Moreover it is clear that producing $\Delta N_\text{eff.}\gtrsim0.06$ only occurs in regions where there is either too much DM or the freeze-in approximation for   DM   production is not applicable because the production rates from relativistic SM fermions thermalize. Moreover we see that for larger allowed DM masses there is actually less $\Delta N_\text{eff.}$. The reason is simply that larger $m_\text{DM}$ at constant $T_\text{RH}$ require larger $v_\text{B-L}$ to fix the relic density, which decreases $\Delta N_\text{eff.}\sim v_\text{B-L}^{-4}$. Consequently our scenario for FIMP DM predicts only a small value of $\Delta N_\text{eff.}$ despite the fact that we introduce two $\nu_R$ and a rather light DM candidate.
\newline
This makes the present construction  different from the cosmology of other (Dirac) neutrino mass models like e.g. the neutrino-philic Two-Higgs-Doublet model  \cite{Wang:2006jy,Gabriel:2006ns,Sher:2011mx,Zhou:2011rc,Davidson:2009ha} or its gauged variations such as \cite{Farzan:2016wym,Farzan:2017xzy,PhysRevD.99.035003,Berbig:2020wve,Bally:2020yid,Li:2022yna} which usually feature light mediators below the EW scale that unavoidably thermalize  the $\nu_R$ and  themselves  leading to $\Delta N_\text{eff.} > \mathcal{O}(0.1)$  \cite{Abazajian:2019oqj,FileviezPerez:2019cyn}. Another interesting scheme is called  \enquote{Common Origin of Warm and Relativistic Decay Products} (COWaRD) \cite{Buen-Abad:2019opj}, where DM and DR are produced together from the decay of a parent particle and the amount of $\Delta N_\text{eff.}$ is correlated with the warmness of DM. There a non-zero $\Delta N_\text{eff.}$ can help to reduce the $\sigma_8$-tension for large scale structures \cite{Douspis:2018xlj,Nunes:2021ipq}. In a sense the COWaRD scheme is the opposite of our idea as it involves thermal DM and predicts a larger amount of DR. All of these models have in common that the more stringent limits on $\Delta N_\text{eff.}$ will already constrain significant amount of their parameter space or even exclude them completely. The only ways to exclude our scenario would be CMB experiments in the far future with a sensitivity to even smaller values of $\Delta N_\text{eff.}=\mathcal{O}(10^{-3})$ or the actual observation of a signal with $0.28>\Delta N_\text{eff.}>0.012$, which by itself would be a smoking gun for different BSM physics.

\section{Inflation and candidates for the inflaton}\label{sec:infl}
\noindent The assumed production mode for DM crucially relies on a low value of the reheating temperature  $\SI{4}{\mega\electronvolt}\lesssim T_\text{RH}\lesssim \SI{5}{\giga\electronvolt}$ together with the assumption of no primordial DM abundance from e.g. inflaton decays during reheating. This puts non-trivial constraints on the explicit  realization of inflation. Of course one can assume that the scalar field responsible for creating the inflationary phase of cosmic expansions is another scalar field with no couplings to the DM. However the present model already contains four different scalar multiplets so a minimal solution is to embed the inflaton into one of them. For concreteness we will assume that the candidate field for inflation is the real component of a complex scalar field $\omega$. Recent Planck constraints \cite{Planck:2018jri} disfavour monomial inflation of the form $\text{Re}(\omega)^p$ with $p>1$ because their potential is too steep leading to a too large tensor to scalar ratio. This is why we only investigate scenarios with a non-minimal coupling of the inflaton to  gravity: This scenario is known as Starobinsky-like inflation   \cite{STAROBINSKY198099,PhysRevD.40.1753,PhysRevD.41.1783,PhysRevD.52.4295,PhysRevD.59.064029,Bezrukov:2007ep,Bezrukov:2008ej,Bezrukov:2008ut} and the action in the Jordan frame reads
\begin{equation}
    \mathcal{S}  =  \int \text{d}^4 x\;\sqrt{-g} \left( \frac{1}{2}M_\text{Pl.}^2+ \xi_\omega \left|\omega\right|^2\right)R.
\end{equation}
In this context we denote the determinant of the metric as $g$, the Ricci curvature scalar as $R$ and $\xi_\omega$ is a dimensionless coupling. On can single out a scalar $\omega$ field to play the role of the inflaton by imposing that the couplings of the other scalar fields satisfy $\lambda_\omega / \xi_\omega^2 \ll \lambda_i / \xi_i^2$ \cite{Choubey:2017hsq}, where the $\lambda$ denote the scalar self couplings. The remaining fields will be treated as spectator fields. We will use the constraints from reheating to find the appropriate inflaton candidate in our model. Due to the presence of additional scalars besides the inflaton there is the possibility of creating isocurvature perturbations as in multi-field inflation models \cite{Kaiser:2013sna,Schutz:2013fua}, which could come into conflict with CMB bounds. Essentially the problem is that massless particles are sensitive to quantum fluctuations during inflation \cite{Beltran:2006sq}. However large isocurvature fluctuations can be prevented if either the tree-level mass or the effective mass generated from inflaton oscillations during reheating is larger than the Hubble rate during inflation \cite{Bettoni:2018utf}. Since both $\eta, \sigma$ have tree-level masses unconnected to any vev and potentially receive effective masses, we do not expect isocurvature perturbations in these directions. Similarly if we assume that the scale of B-L breaking $v_\text{B-L}$ is larger than the Hubble rate $H_I$ during inflation and $\text{U}(1)_\text{B-L}$ is never restored, then the would-be-Goldstone mode $\varphi_I$ corresponds to the longitudinal mode of the massive $Z'$ and not to a  massless field. Reference  \cite{Choubey:2017hsq} found that in the extension of the SM with an inert doublet $\eta$ housing the inflaton there are only negligible isocurvature fluctuations. A detailed investigation of these fluctuations for the full model is beyond the scope of the present study and we will be content with just outlining how inflation could be realized.
\newline
Note that we can also allow for a temperature at the end of inflation far above the MeV and GeV range if there is an additional long lived particle that dominates the energy budget of the universe. This leads to an intermediate matter dominated phase \cite{PhysRevD.31.681} which can end in a second radiation dominated epoch with a smaller temperature of the required order of magnitude.

\subsection{The SM like Higgs} 
\noindent Using the SM like Higgs as the inflaton \cite{Bezrukov:2007ep,Bezrukov:2008ej,Bezrukov:2008ut,Bezrukov:2009db,Garcia-Bellido:2011kqb,Barvinsky:2008ia,Barvinsky:2009fy} is a very minimal scenario see \cite{Rubio:2018ogq} for a review.
The main drawback of this approach is that the measured value of the Higgs self coupling $\lambda_H$ requires a rather large value of $\xi_H = \mathcal{O}\left(10^4\right)$, which might give rise to unitarity problems \cite{Lerner:2009na,Burgess:2010zq,Hertzberg:2010dc} at scales above $M_\text{Pl}/ \xi_H$. The unitarity problem could for instance be cured by
assuming a different coupling to gravity \cite{Escriva:2016cwl,Fumagalli:2017cdo,Tenkanen:2019jiq}. Another possibility is exploiting that the SM Higgs self coupling $\lambda_H$ becomes very small at large energy scales which flattens the potential and leads to $\xi = \mathcal{O}\left(10\right)$, which is known as critical Higgs inflation \cite{Hamada:2013mya,Hamada:2014iga,Bezrukov:2014bra,Hamada:2014wna,Drees:2019xpp}.
In terms of BSM physics there is also the attractive possibility to invoke additional scalars to modify the Higgs potential see e.g. \cite{Giudice:2010ka,Kahlhoefer:2015jma,Ballesteros:2016xej}. While it would be interesting to see whether the additional scalars present in this model can solve the unitarity problem it would definitely require a dedicated analysis beyond this work. Consequently we do not consider Higgs inflation further and investigate the other scalar fields as inflaton candidates.

\subsection{The B-L breaking singlet}
\noindent 
The only singlet with a B-L breaking vev could be the inflaton too \cite{Okada:2011en,Buchmuller:2012wn}. We neglect the mixing between $h$ and $\varphi$ because the EW gauge symmetry is restored at large temperatures \cite{Kirzhnits:1972iw,KIRZHNITS1972471} so the mixing term vanishes together with $v_H$. For the same reason we compute the decays to the entire doublet $H$ and not just $h$. For the purpose of finding estimates we work in the regime of perturbative reheating. We assume that all additional scalars, fermions and the $Z'$ are heavier than the inflaton so the only available decay modes are 
\begin{equation}\label{eq:dec-rh}
    \Gamma\left(\varphi\rightarrow H^\dagger H\right)= \frac{\lambda_{H\phi}^2 v_\text{B-L}^2}{8 \pi m_\varphi}, \quad \text{and} \quad \Gamma\left(\varphi\rightarrow \overline{\chi}\chi\right) = \left(\frac{m_\text{DM}}{v_\text{B-L}}\right)^2 \frac{m_\varphi}{8\pi},
\end{equation}
where the decay width to DM is obtained from \eqref{eq:falpha} after converting it to $\sim m_\text{DM}/v_H$ and replacing $v_H \rightarrow v_\text{B-L}$. Since the scalar will oscillate in its potential during reheating it develops an effective mass depending on its oscillation frequency and the same goes for all other scalar  fields as well as the $Z'$ since they share quartic couplings with $\varphi$. Hence requiring that \eqref{eq:dec-rh} are the only available decay modes and that e.g. $\varphi\rightarrow S_1 S_2 h$ is absent amounts to a bound on the effective field dependent masses and not on the tree-level masses that we have employed so far. For the sake of simplicity this first estimate will work exclusively with the tree level masses. 
If we want to avoid a primordial abundance of DM the first step is to make sure that decays to SM particles dominate the reheating process 
\begin{equation}
    \text{BR}\equiv \frac{\Gamma\left(\varphi\rightarrow \overline{\chi}\chi\right)}{\Gamma\left(\varphi\rightarrow \overline{\chi}\chi\right)+\Gamma\left(\varphi\rightarrow H^\dagger H\right)}\simeq \frac{\Gamma\left(\varphi\rightarrow \overline{\chi}\chi\right)}{\Gamma\left(\varphi\rightarrow H^\dagger H\right)}= \frac{1}{\lambda_{H\phi}^2} \frac{m_\text{DM}^2}{v_\text{B-L}^2} \frac{m_\varphi^2}{v_\text{B-L}^2}\ll 1,
\end{equation}
which sets bounds on the model parameters. Assuming BR$\;\ll1$ we can determine the reheating temperature from the decay to the SM Higgses, which themselves will decay to fermions creating a hot thermal bath. In this limit the reheating temperature is found to be
\begin{equation}\label{eq:TRHdef}
    T_\text{RH} = \sqrt{\frac{2}{\pi}} \left(\frac{10}{g_{*\rho}(T_\text{RH})}\right)\sqrt{M_\text{Pl} \Gamma\left(\varphi\rightarrow H^\dagger H\right)}.
\end{equation}
The assumed range of reheating temperatures for DM production requires that either $\lambda_{H\phi}\ll 1$ or that $m_\varphi\gg v_\text{B-L}$. However the second condition can not be realized because $m_\varphi$ is proportional to $v_\text{B-L}$ according to \eqref{eq:scal-masses} and we can not make $m_\varphi$ arbitrarily heavy due to the perturbativity limit $\lambda_\phi<\sqrt{4\pi}$.\\
Inflaton decays can produce DM as well and the corresponding Boltzmann equation during reheating reads \cite{Bernal:2021qrl}
\begin{equation}
    \frac{\text{d}n_\chi}{\text{d}t}+ 3 H n_\chi=  \frac{\rho_\varphi}{m_\varphi} \Gamma\left(\varphi\rightarrow \overline{\chi}\chi\right), \quad \text{with} \quad H^2 = \frac{\rho_\varphi+\rho_\text{SM}}{3 M_\text{Pl}^2}
\end{equation}
and we denote the energy density of the non-relativistic inflaton condensate as  $\rho_\varphi$.
The DM yield today is found to be \cite{Bernal:2021qrl}
\begin{equation}
    Y_\chi(T_0) \simeq\frac{3}{4} \frac{g_{*\rho}(T_\text{RH})}{g_{*S}(T_\text{RH})} \frac{T_\text{RH}}{m_\varphi}\text{BR}
\end{equation}
and the DM energy density today can be calculated with \eqref{eq:conv}. 
For simplicity we assume $g_{*\rho}(T_\text{RH})\simeq g_{*S}(T_\text{RH}) $.
We trade the inflaton mass via equation \eqref{eq:TRHdef} for an expression involving $T_\text{RH}$ and $v_\text{B-L}$, where the dependence on $\lambda_{H\phi}$ in $ Y_\chi(T_0)$ divides out.  By using our limit on $v_\text{B-L}$ in \eqref{eq:vev-cosmo} we derive an upper-limit on the relic abundance from inflaton decays 
\begin{equation}
     \Omega_\text{DM}^\text{inf.} h^2 \lesssim 0.56\; \left(\frac{m_\text{DM}}{\SI{10}{\kilo\electronvolt}}\right)^3 \cdot \left(\frac{\SI{1}{\giga\electronvolt} }{T_\text{RH}}\right)^\frac{5}{2} \cdot \sqrt{\frac{11.67}{\sum_i N_i(T_\text{RH})}} \cdot  \left(\frac{76}{g_{*\rho}(T_\text{RH})}\right)^\frac{7}{4}.
\end{equation}
It is evident that large $m_\text{DM}$ and low reheating temperatures could lead to an abundance that is larger than the FIMP one in \eqref{eq:relic-abun}. Demanding that the abundance from inflaton decays does not overclose the universe cuts away all the available parameter space in \ref{fig:paramspaceGauge}.
There is another reason why this channel is not suited for light DM production: Since the production mode is different from both freeze-in (which requires a thermal bath) and thermal production, the phase space distribution and hence the velocity distribution of the DM will be different assuming all of DM was produced via this single channel. This manifest itself in a modified Lymann-$\alpha$ bound \cite{Masina:2020xhk,Bernal:2021qrl}
\begin{equation}
    m_\text{DM}\gtrsim \SI{2}{\kilo\electronvolt} \cdot \left(\frac{m_\varphi}{T_\text{RH}}\right)\cdot\left(\frac{m_\text{WDM}}{\SI{3.5}{\kilo\electronvolt}}\right),
\end{equation}
which was recast from the bound for thermally produced DM with $m_\text{WDM}\gtrsim\SI{3.5}{\kilo\electronvolt}$ \cite{Irsic:2017ixq} (which is the average of the two possible warm DM masses in section \ref{sec:lyman}). If we assume that $m_\varphi =\mathcal{O}\left( v_\text{B-L}\right)$ then we expect an inflaton with at least a TeV scale mass (see \eqref{eq:vev-lab}), which is much larger than the assumed MeV-GeV reheating temperatures. Therefore the DM mass for inflaton production would be orders of magnitude larger than 2 keV and potentially violates the bound from invisible Higgs decays in \eqref{eq:DMBound1} and could lead to overclosure. Thus we conclude that for our purposes  $\varphi$ can not be the inflaton, because it tends to produce too much DM. Therefore we assume that $\varphi$ is too heavy to be produced during reheating.

\subsection{The inert doublet or singlet scalars}\label{sec:inert-inf}
\noindent 
As previously mentioned $v_H$ vanishes due to the restoration of the EW symmetry at large temperatures  \cite{Kirzhnits:1972iw,KIRZHNITS1972471} .
In this limit   we can  relate $S_1=\eta_R^0$ as well as $S_2= \sigma_R^0$  and consider each field as a candidate individually. Similar to Higgs inflation the inert doublet $\eta$ can house the inflaton \cite{Choubey:2017hsq}. This scenario is free of the unitarity problem because the   value of the $\eta$ self coupling $\lambda_\eta$ is unconstrained by phenomenology.  We can not just reuse the perturbative reheating estimate \eqref{eq:TRHdef} from the previous section, because without a vev  there is no tree level decay to Higgses like in \eqref{eq:dec-rh} or to EW gauge bosons for $\eta_R^0$.
In this model reheating occurs via quartic couplings to electroweak gauge bosons and SM Higgses \cite{Bezrukov:2008ut,Choubey:2017hsq} and we assume that the $Z'$ is too heavy to be produced. Reheating typically takes place through resonant gauge boson production which then annihilate to SM fermions. In this scenario the reheating temperature was found to be \cite{Choubey:2017hsq}
\begin{equation}\label{eq:TRH-gaug}
    T_\text{RH}^\eta \simeq 10^{14}\;\text{GeV}\;\lambda_\eta^{-\frac{1}{8}}.
\end{equation}
Generating sub-GeV reheating temperatures is impossible in this regime, because it would require non-perturbative values of $\lambda_\eta$. We conclude that another reheating channel is needed and hence consider an inflaton without SM gauge interactions:
$\sigma$  is an SM singlet and has no vev as well. If we assume that the effective field dependent mass of the $Z'$ is too large to be produced then creating  SM Higgses via the quartic coupling $\lambda_{H\sigma}$ in \eqref{eq:Hsigma} is the only possibility left. Since this  process depends on the new coupling $\lambda_{H\sigma}$ instead of the known SM gauge couplings the reheating temperature will also depend on this  unconstrained parameter. Subsequent decays and annihilations of the Higgs to SM states then seed the SM radiation bath.
Reference \cite{Lerner:2011ge} found that for resonant Higgs production
\begin{equation}
    T_\text{RH}^{\sigma\;\text{res.}}\simeq \SI{3e13}{\giga\electronvolt}\; \left(\frac{\lambda_\sigma}{\lambda_{H\sigma}^2}\right)^\frac{1}{4}.
\end{equation}
The analysis \cite{Lerner:2011ge} made the conservative assumption of having reheating occur during the quadratic phase of the potential before the quartic self-interaction of the inflaton becomes dominant, which can be expressed as $\lambda_\sigma > 0.25\; \lambda_{H\sigma}$ \cite{Lerner:2011ge}. If we drop this assumption, which  \cite{Lerner:2011ge} emphasizes is not ruled out, we can choose smaller values of $\lambda_\sigma\ll \lambda_{H\sigma}$  and   can at least  in principle accommodate the range $\SI{4}{\mega\electronvolt}\lesssim T_\text{RH}\lesssim \SI{5}{\giga\electronvolt}$. The authors of \cite{Lerner:2011ge} also found that reheating can  occur in another regime if inflaton excitations annihilate into pairs of Higgs bosons leading to the estimate
\begin{equation}
    T_\text{RH}^{\sigma\;\text{ann.}}\simeq \SI{9e13}{\giga\electronvolt}\cdot \lambda_\sigma^\frac{1}{4}.
\end{equation}
The conservative assumption about reheating occurring in the quadratic regime of the potential would lead to $\lambda_\sigma>0.019$ \cite{Lerner:2011ge}, but again we need to drop this assumption and require $\lambda_\sigma\ll1$ to obtain the phenomenologically favoured reheating temperatures. In the next section \ref{sec:baryo} we will introduce a decay of $\sigma$ to exotic quarks, which might open up another possibility for realizing the required reheating temperature.
\newline 
Let us emphasize that there are bounds from vacuum stability and perturbativity on the quartic couplings 
\cite{Lerner:2009xg}, but since these bounds are usually obtained in models with a simpler scalar sector it requires a dedicated study to translate them to our construction, because of e.g. threshold effects from heavy scalars \cite{Elias-Miro:2012eoi}. Note that at some point during reheating there will be the SSB of the EW symmetry generating a coupling of  $\sigma_R^0$ to the EW gauge bosons proportional to $\sin(\alpha)$.
But since the neutrino mass  \eqref{eq:act-mass} does not directly depend on the mixing angle $\alpha$ in the radiative seesaw limit  we can make this mixing small.
\newline
If we assume that the $F$ fermions are heavier than the $\sigma$ there will be no inflaton decays to $\chi$ via the Yukawa interaction in \eqref{eq:BSMYuk2}. The only way to generate the unwanted primordial DM population would be annihilation processes of the form $\sigma_R^0 \sigma_R^0 \rightarrow \overline{\chi}\chi$ mediated by heavy $F$s. We do not expect this to lead to a significant DM abundance, because scattering is inefficient for non-relativistic excitations of the inflaton field and the production is suppressed by the heavy $F$ mass. On top of that the DM production competes with the unsuppressed process for creating the SM radiation $\sigma_R^0 \sigma_R^0 \rightarrow H^\dagger H$. Since the singlet scalar might not have decay modes, we need to ensure that the inflaton becomes a subdominant component of the universe's energy budget after reheating. The additional interactions like Higgs or $Z'$ mediated scatterings with the SM fermions  could help thermalize the inflaton with the radiation bath, which is already in thermal equilibrium  \cite{Harigaya:2013vwa}.
We conclude that the only possible inflaton candidate that is not in conflict with the cosmological DM and reheating requirements is $\sigma_R^0$.

\section{Baryogenesis}\label{sec:baryo}
\noindent The assumed low scale reheating is hard to reconcile with most known mechanisms \cite{Yoshimura:1979gy,Weinberg:1979bt,Fry:1980bd} for Baryogenesis. Leptogenesis \cite{Fukugita:1986hr} for instance relies on producing a leptonic asymmetry that gets converted into a baryon asymmetry by electroweak sphaleron processes, which are in equilibrium only above the EW phase transition at $T_\text{EW}=\mathcal{O}(\SI{100}{\giga\electronvolt})$. On top of that since the SM neutrinos do not mix with any of the heavy new neutrinos $N,F$ we can not realize leptogenesis via oscillations \cite{Akhmedov:1998qx} as well. Thus we are left with mechanisms that do not rely on the sphaleron transition above the EW scale. One example is the spontaneous Baryogenesis \cite{COHEN1987251,Cohen:1988kt} mechanism, which however needs reheating temperatures far above the assumed MeV-GeV scale window. Hence some other form of non-thermal baryogenesis during reheating seems to be the only possibility left if we insist that the temperature at the end inflation is indeed in the previously mentioned range.
\newline
The Affleck-Dine mechanism \cite{AFFLECK1985361} relies on baryon number charged scalars whose real and imaginary parts evolve non-trivially in time, which acts as a source term for baryon number. This scenario can in principle operate at low reheating temperatures if the initial field value of the Affleck-Dine field is very large compared to its mass. Since all of our scalars except $H$ are charged under B-L this is an attractive possibility. For concreteness we will treat $\sigma$ as the Affleck-Dine field; whether it can accommodate both baryogenesis and inflation at the same time like e.g. \cite{Charng:2008ke,Hertzberg:2013jba,Hertzberg:2013mba,Takeda:2014eoa,Bettoni:2018utf,Cline:2019fxx,Cline:2020mdt,Lin:2020lmr,Lloyd-Stubbs:2020sed,Barrie:2021mwi,Mohapatra:2021ozu,Mohapatra:2022ngo} will be left for future investigation.
An important ingredient is a small explicit Baryon number breaking interaction. Of course we can not break our gauged B-L explicitly but a term of the form $\lambda_\text{AD} (\sigma^4+\sigma^{*\;4})$ could arise after the spontaneous breaking of $\text{U}(1)_\text{B-L}$. To do so we allow for the small $\mathcal{Z}_5$ breaking term 
\begin{equation}\label{eq:ADoperator}
    \mathcal{L} \subset -\lambda'\left(\sigma^2 \phi^2+\text{h.c.}\right),
\end{equation}
which after integrating out the heavy radial mode $\varphi$ (we ignore the $\varphi$-Higgs mixing here) leads to an  operator
\begin{equation}\label{eq:opAD}
     \mathcal{L}_\text{EFT} \subset -\frac{\lambda'^{\;2} v_\text{B-L}^2}{m_\varphi^2} \left(\sigma^4+\sigma^{*\;4}\right)
\end{equation}
and we can identify $\lambda_\text{AD} =  \lambda'^{\;2} v_\text{B-L}^2 / m_\varphi^2\simeq \lambda'^{\;2} / (2\; \lambda_\phi) $ from \eqref{eq:scal-masses}. Quite interestingly this allows us to make $\lambda_\text{AD}$ small just by assuming $\sqrt{\lambda'}\ll  \lambda_\phi $. Since $\lambda'\rightarrow 0$ would restore the discrete symmetry the choice $\lambda'\ll1$ is technically natural \cite{tHooft:1979rat}. Of course assuming the existence of this operator begs the question why the other $\mathcal{Z}_5$ breaking interactions are absent. 
The last missing ingredient is a way to transmit the $\sigma$-asymmetry to the quarks. To do so we introduce a pair of heavy  vector-like quarks $\left(Q_L,Q_R\right)$ that are weak isospin singlets with the hypercharge  $Y=-2/3\;\left(4/3\right)$ of the right chiral down (up) quarks. The quarks come with a B-L charge $Q_\sigma +1/3=-2/3$ and transform  as $\omega^{-4}$ under $\mathcal{Z}_5$, where $\omega = e^\frac{2 i\pi}{5}$, so that we can realize the operators
\begin{equation}
    \mathcal{L}\subset - Y_{Qq} \overline{Q_L}  \sigma d_R -m_Q \overline{Q_L}Q_R\;.
\end{equation}
Here $d_R$ can in principle also be replaced with $u_R$; we chose the hypercharge $-2/3$ to make the vector-like quarks resemble the down-type quarks which might help with unification \cite{GURSEY1976177,SHAFI1978301}. The above interaction could also lead to inflaton mediated washout scatterings depleting the baryon asymmetry \cite{Lloyd-Stubbs:2020sed}, which puts constraints on the coupling $ Y_{Qq}$.
In order to prevent stable exotic quarks from forming relics \cite{stablequarks} we have to   demand that $m_Q > m_\sigma $ so the  $Q$ can decay via the above operator to $\sigma \;u_R$ in the late universe.  Alternatively one can also arrange for $m_\sigma>m_Q > 2 m_h$ instead  so that the decay of the vector-like quarks proceeds via off-shell $\sigma$ as $Q_L\rightarrow \sigma^* + d_R \rightarrow 2 h + d_R$; the Higgses then further decay to SM states. In the early universe the field $\sigma$ receives a potentially large effective mass from inflaton oscillations during reheating so for both aforementioned cases the $CP$-conserving decay $\sigma \rightarrow \overline{Q_L}u_R$ would be possible and one can indeed transmit the asymmetry from the Affleck-Dine field to the quark sector. This decay could open up another interesting reheating scenario as well. In the following we will assume that $T_\text{RH}$ arises either from to the channels enumerated in the previous section \ref{sec:inert-inf} or via the aforementioned decay. An estimate for the baryon asymmetry leads to \cite{Harigaya:2016hqz,Harigaya:2019emn}
\begin{equation}
    \frac{n_\text{B}}{s}\simeq 10^{-10}\;\cdot \left(\frac{\lambda_\text{AD}}{10^{-2}}\right) \cdot \left(\frac{\sin\left(4\theta_i\right)}{0.5}\right)\cdot \left(\frac{r_i/m_\sigma}{6\times10^6}\right)^3 \cdot \left( \frac{r_i}{6\times10^{9}\;\text{GeV}}\right) \cdot \left(\frac{T_\text{RH}}{\SI{1}{\giga\electronvolt}}\right).
\end{equation}
Here we use the polar parameterization for $\sigma$, where $r_i$ is the initial value of the radial component  and $\theta_i$ denotes the initial angle needed for $CP$ violation. This decomposition should not to be confused with the cartesian representation from \eqref{eq:reps}.
The initial angle can not be set to  arbitrarily small values in order to avoid  isocurvature perturbations \cite{Barrie:2021mwi}, which is why we chose $\sin\left(4\theta_i\right)\simeq 0.5$.
We see that very large initial field values are needed to compensate for the low reheating temperature. Such a high field value of $r_i/m_\sigma\simeq 6\times 10^6$ usually requires a very flat potential and could be an initial condition. Alternatively the non-minimal coupling to gravity might help  to generate this field value dynamically \cite{Dine:1995kz}:
It was found that this coupling  together with the tree level mass squared creates an effective mass squared depending on the Hubble parameter. This effective mass is tachyonic at early times when $H\gg m_\sigma$ and later turns real again, which can be understood as an inverted phase transition \cite{Weinberg:1974hy,PhysRevLett.45.1}. Afterwards the field, which can be visualized as an over-damped oscillator, is stuck in its previous non-trivial minimum corresponding to an initial value of \cite{Rubakov:2017xzr}
\begin{equation}
    r_i \simeq \sqrt{\frac{\xi_\sigma}{\lambda_\sigma}}m_\sigma,
\end{equation}
before it starts to relax to its true minimum $\sigma=0$ as soon as  the Hubble rate satisfies $H\sim m_\sigma$ provided that $\lambda_\text{AD}\ll \lambda_\sigma$. From this mechanism we can deduce that a scalar self coupling of 
\begin{equation}
    \lambda_\sigma \simeq   2.8\times10^{-14} \cdot \xi_\sigma \cdot \left( \frac{r_i}{6\times10^{9}\;\text{GeV}}\right)^2 \cdot \left(\frac{1\;\text{TeV}}{m_\sigma}\right)^2
\end{equation}
would be required for the initial field value and a scalar mass in accord with our previous estimates \eqref{eq:numass} and \eqref{eq:mDM}. Note that this violates the previous assumption $\lambda_\text{AD}\ll \lambda_\sigma$, but we can reconcile this by assuming that the heavy $\varphi$ will only be integrated out at temperatures somewhat below the inverted phase transition so that the operator \eqref{eq:opAD} is absent initially. On the level of estimates it seems that our scalar potential can reproduce the observed baryon asymmetry, but again we stress that it requires a separate study to work out the details especially in the inflationary context and considering the radiative stability.
\newline
It is noteworthy that the operator \eqref{eq:ADoperator} also sources a mass splitting $\simeq \pm\lambda' v_\text{B-L}^2$ between the real and imaginary parts of $\sigma$, while for the neutrino mass generation we assumed that they are mass degenerate. Under the assumption that this additional mass splitting is small compared to the overall mass scale of the $S_{1,2}(A_{1,2})$ and the mass splitting between the different generations of scalars our conclusions about the neutrino and DM masses are unchanged.
\newline
If $T_\text{RH}$ is the temperature after an intermediate epoch of matter domination and the true temperature at the beginning of the first radiation dominated phase was far above the electroweak scale this allows for the other previously discussed mechanisms again. In that case the challenge is to generate enough entropy to dilute unwanted relics (such as thermally produced DM) while retaining enough baryon asymmetry \cite{Hasenkamp:2010if}.

\section{Summary}
\noindent We presented an extension of the Dirac scotogenic model \cite{Gu:2007ug,Farzan:2012sa} that creates the Dirac mass of a light fermionic DM candidate $\chi$ together with the active neutrino masses via one-loop diagrams. The model relied on a gauged $\text{U}(1)_\text{B-L}$ symmetry, whose anomaly-freedom determined the charges of the DM and two copies of $\nu_R$. We found that our symmetry based approach predicts that only two SM neutrinos are massive Dirac fermions, whereas the third one remains exactly massless, because there is no third $\nu_R$.
In order to ensure the DM stability and to prevent unwanted operators that could affect the neutrino or DM mass generation we had to impose a separate $\mathcal{Z}_5$ symmetry as well.
Additionally one requires an inert scalar doublet $\eta$ and an inert singlet $\sigma$ together with the B-L breaking scalar singlet $\phi$. Moreover we had to introduce a host of vector-like fermions to generate the necessary loop diagrams. It was found that the vector-like leptons $F$ needed for the DM masses couple to the SM Higgs and are light enough to potentially be probed by next generation collider experiments.
\newline
We then chose a minimal scenario where we assumed that only the SM degrees of freedom augmented by two $\nu_R$ and $\chi$ are present after reheating.
The constraint from invisible Higgs decays enforces $m_\text{DM} \lesssim \SI{2}{\giga\electronvolt}$
and  the DM mass has to be larger than $ (4-16)\;\text{keV}$ due to the Lyman-$\alpha$ forest. After demonstrating that thermal production and out of equilibrium Higgs decays both lead to an over-production of DM, we were able to narrow the window of the allowed reheating temperatures down to the range between about $\SI{5}{\giga\electronvolt}$ and $\SI{4}{\mega\electronvolt}$. Consequently we analyzed the joint production of DM $\chi$ and DR $\nu_R$ from out of equilibrium annihilations of the SM fermions via the B-L gauge boson $Z'$. The DM mass has to be smaller than  $\mathcal{O}(\text{MeV})$ in order to suppress DM production via diagrams with an intermediate SM like Higgs compared to $Z'$ mediated scatterings. We found a potentially viable parameter space with $v_\text{B-L}\gtrsim\mathcal{O}\left(\SI{10}{\tera\electronvolt}\right)$ that leads to the correct observed DM abundance but predicts  $\Delta N_\text{eff.}\lesssim 0.012$. The amount of produced dark radiation decreases with the DM mass so in a sense $m_\text{DM}$ and $\Delta N_\text{eff.}$ are anti-correlated. 
This is in striking contrast to other Dirac neutrino and DM mediator models which usually predict larger  $\Delta N_\text{eff.}$. Thus while the aforementioned models can already be tested or ruled out by tightening the observational bounds on $\Delta N_\text{eff.}$, only the  detection of $\Delta N_\text{eff.}> 0.012$ could falsify our DM production scenario in the near future.
\newline
Owing to the fact that we need a very low reheating temperature and want a negligible primordial DM abundance we were able to single out the real component of the $\sigma$ field to play the role of the inflaton.
In addition we found a way for how the $\sigma$ field can also potentially realize Affleck-Dine baryogenesis if we introduce  a small source of $\mathcal{Z}_5$-breaking in the scalar potential together with a pair of vector-like down quarks.  We leave a detailed study of the inflationary predictions, reheating and non-thermal baryogenesis  for future investigation.
\newline
To summarize, we introduced a new abelian gauge theory that can simultaneously explain the active neutrino and  fermionic dark matter  masses via loop diagrams. Our construction produces the observed DM relic abundance together with minuscule amounts of dark radiation in the freeze-in regime and can potentially account for inflation, reheating and Baryogenesis.

\section*{Acknowledgments}
\noindent 
This work benefited from the use of \verb|PackageX| \cite{Patel:2015tea,Patel:2016fam}.
We would like to thank  Nicolás Bernal and Andreas Trautner for helpful comments about the manuscript as well as Rahul Mehra for insightful discussions about DM direct detection using electron targets.

\bibliographystyle{JHEP}
\bibliography{references}

\end{document}